\documentclass[a4paper,11pt]{article} 
\pdfoutput=1 
\usepackage{jheppub} 

\usepackage{amsmath,graphicx,hyperref}
\usepackage{xcolor}
\usepackage{enumitem}

\allowdisplaybreaks

\newcommand{\fms}[1]{{#1}\!\!\!/}

\newcommand{\mc}{\mathcal}
\newcommand{\mr}{\mathrm}
\newcommand{\mP}{\mathcal{P}}
\newcommand{\mO}{\mathcal{O}}
\newcommand{\mQ}{\mathcal{Q}}
\newcommand{\mD}{\mathcal{D}}

\newcommand{\be}{\begin{equation}} 
\newcommand{\ee}{\end{equation}} 
\newcommand{\bea}{\begin{eqnarray}} 
\newcommand{\eea}{\end{eqnarray}}

\newcommand{\n}{\overline{n}}
\newcommand{\nn}{\frac{\fms{\overline{n}}}{2}}

\newcommand{\bl}[1]{{\bf{#1}}}

\newcommand{\blp}[1]{{\bf{#1}}_{\perp}}
\newcommand{\blpu}[1]{{\bf{#1}}^{\perp}}

\newcommand{\bsp}[1]{{\boldsymbol{#1}}_{\perp}}

\newcommand{\nnb}{\nonumber} 

\newcommand{\as}{\alpha_s} 

\newcommand{\eps}{\epsilon} 
\newcommand{\veps}{\varepsilon} 

\newcommand{\euv}{\eps_{\mr{UV}}}
\newcommand{\eir}{\eps_{\mr{IR}}}

\newcommand{\pp}{\blp{p}}
\newcommand{\qp}{\blp{q}}
\newcommand{\kp}{\blp{k}}
\newcommand{\lp}{\blp{l}}
\newcommand{\pps}{\blp{p}^2}
\newcommand{\qps}{\blp{q}^2}
\newcommand{\kps}{\blp{k}^2}
\newcommand{\lps}{\blp{l}^2}

\newcommand{\bb}{\bl{b}}

\begin{document}

%\vspace*{18pt}

%%%%%%%%%%%%%%%%%%%%%%%%%%%%%%%%%%%%%%%%%%%%%%%%%%%%%%%%%%%%%%%%%%%%%%
%%%%%%%%%%%%%%%%%%%%%%%%%%%%% Title %%%%%%%%%%%%%%%%%%%%%%%%%%%%%%%%%%
%%%%%%%%%%%%%%%%%%%%%%%%%%%%%%%%%%%%%%%%%%%%%%%%%%%%%%%%%%%%%%%%%%%%%%

\title{Heavy quark transverse momentum dependent fragmentation}

\def\Seoultech{Institute of Convergence Fundamental Studies and School of Natural Sciences, Seoul National University of Science and Technology, Seoul 01811, Korea}
\def\Pitt{Pittsburgh Particle Physics Astrophysics and Cosmology Center (PITT PACC) \\ Department of Physics and Astronomy, University of Pittsburgh, Pittsburgh, Pennsylvania 15260, USA}
\def\TUMPhys{Physik Department, Technische Universit\"{a}t M\"{u}nchen, James-Franck-Str. 1, 85748 Garching, Germany}

\author[a]{Lin Dai}
\emailAdd{lin.dai@tum.de}
\affiliation[a]{\TUMPhys}
\author[b]{Chul Kim}
\emailAdd{chul@seoultech.ac.kr}
\affiliation[b]{\Seoultech} 
\author[c]{Adam K. Leibovich}
\emailAdd{akl2@pitt.edu}
\affiliation[c]{\Pitt}  

\preprint{TUM-EFT 177/23}

\abstract{ %\vspace{0.1cm}\baselineskip 3.0ex 
In this paper, we investigate the heavy quark (HQ) mass effects on the transverse momentum dependent fragmentation function (TMDFF). We first calculate the one-loop TMDFF initiated by a heavy quark. We then investigate the HQ TMDFF in the limit where the transverse momentum, $q_\perp$ is small compared to the heavy quark mass, $q_\perp \ll m$ and  also in the opposite limit where $q_\perp \gg m$. As applications of the HQ TMDFF, we study the HQ transverse momentum dependent jet fragmentation function, where the heavy quark fragments into a jet containing a heavy hadron, and we investigate a heavy hadron's transverse momentum dependent distribution with respect to the thrust axis in $e^+e^-$ collisions.

}

\maketitle 

%%%%%%%%%%%%%%%%%%%%%%%%%%%%%%%%%%%%%%%%%%%%%%%%%%%%%%%%%%%%%%%%%%%%%%
%\baselineskip 3.5ex 

\section{Introduction} 

Much work has been done recently investigating transverse momentum dependence in high energy scattering processes (see Ref.~\cite{Boussarie:2023izj} for a recent comprehensive review of transverse momentum dependent parton distribution functions and fragmentation functions). The small transverse momentum dependent (TMD) fragmentation function (FF) to a hadron~\cite{Collins:1981uw,Collins:1992kk} is a crucial element in understanding the high energy mechanism of hadronization, providing a three-dimensional picture of the fragmenting process.  A detailed study of the TMDFF can play a decisive role in extracting  precise information on the TMD parton distribution functions (PDFs) in collisions, for instance by a  precise study of the  semi-inclusive deep inelastic scattering. For a detailed review on the TMDFF, we refer the reader to Ref.~\cite{Metz:2016swz,Boussarie:2023izj}. 

Recently, without many nonperturbative inputs, rather clean measurements for TMDFFs have been obtained through jet observations, for example, by measuring the momentum of a hadron within a jet with the reference to the jet axis~\cite{Bain:2016rrv,Neill:2016vbi,Kang:2017glf} or the thrust axis~\cite{Boglione:2017jlh,Belle:2019ywy,Soleymaninia:2019jqo,Boglione:2020auc,Kang:2020yqw,Makris:2020ltr,Gamberg:2021iat,Boglione:2022nzq,DAlesio:2022brl,Boglione:2023duo}. 
While these processes introduce nonglobal logarithms~\cite{Dasgupta:2001sh,Banfi:2002hw}, in the framework of QCD factorization on the jet cross section, it is rather easy to pick up the TMD fragmentation component, which can then be applied to other processes, like the  semi-inclusive deep inelastic scattering mentioned above.  
In addition, recent developments in the treatment of  large rapidity logarithms~\cite{Collins:2011zzd,Aybat:2011zv,Chiu:2011qc,Chiu:2012ir,Echevarria:2012js} make it easier to compare the TMD components of different processes with disparate rapidity gaps~\cite{Ebert:2019okf,Kang:2020yqw}. 

Given that an energetic heavy quark is often produced in high energy collisions, we can also consider the heavy quark (HQ) TMDFF to a heavy hadron, like a $B$ meson, as an extension of the study for a light quark-initiated TMDFF~\cite{Makris:2018npl,delCastillo:2020omr,delCastillo:2021znl,vonKuk:2023jfd,Copeland:2023wbu,Copeland:2023qed}. An interesting feature of the HQ TMDFF is that the heavy quark mass introduces a new scale other than the transverse momentum $\blp{q}$, which complicates the factorization structure of the fragmentation and provides a unique perspective that is distinguishable from the case of a light quark. 

In order to consider various hierarchies between $\blp{q}$ and the heavy quark mass $m$, we need to investigate different factorizations for each kinematic situation, which enable us to systematically resum the large logarithms induced from the large scale separations between $\blp{q}$, $m$, and $Q$, where $Q$ is a typical hard scale comparable to an energy of the boosted heavy quark. 
Furthermore, based on the factorization theorem, we can consider the appropriate parameterization of nonperturbative inputs for the hadronization of the heavy quark. 
In this paper, employing soft-collinear effective theory (SCET)~\cite{Bauer:2000ew,Bauer:2000yr,Bauer:2001yt,Bauer:2002nz},
we construct the factorization theorem of the heavy quark TMDFF, perform next-to-leading order (NLO) calculations on each factorization ingredient, and consider resummation of the large logarithms of $Q$, $\blp{q}$, and $m$. 

The paper is organized as follows. In Section \ref{oneloopcalc}, we calculate the HQ TMDFF at one loop. In Section \ref{smallq}, we investigate the HQ TMDFF in the region of parameter space where $\blp{q}\ll m$, while in Section \ref{TMDFFqmlm}, we look at the other limit, $\blp{q}\gg m$. In Section \ref{TMDFFqgglambda}, we investigate the nonperturbative contributions to  when $\blp{q}\gg \Lambda_{\rm QCD}$.
In Section \ref{HQTMDJFF} we apply the previous results to the case where the initiating heavy quark fragments into a jet containing a heavy meson, by introducing the heavy quark TMD jet fragmentation function. As another application, in Section \ref{thrust} we study the heavy hadron's TMD distribution with respect to the thrust axis in $e^+e^-$ annihilation. We conclude in Section \ref{conclusions}. We also include a few Appendices with some extra information about the calculations.

\section{One loop calculation of the heavy quark TMD fragmentation function} \label{oneloopcalc}

In this section, for a boosted heavy quark, we consider the one loop contribution to the TMD distribution in momentum space without specifying the hierarchy between the transverse momentum $\blp{q}$ and the heavy quark mass $m$ (i.e., $\blp{q} \sim m$). 
Through the calculation in momentum space, which is more intuitive than the calculation in coordinate space, we separate the ultraviolet (UV), the infrared (IR), and the rapidity divergences explicitly. Finally we will show that the one loop result of the heavy quark TMD fragmentation function (TMDFF) is IR-safe and shares the same renormalization behavior for the UV and the rapidity divergences  compared to the case of a light quark.  

In the hadron frame where the transverse momentum of the final observed hadron is set to zero, the heavy quark TMD fragmentation function (TMDFF) is given in $D$ dimensions by~\cite{Collins:1981uw} 
\be 
\label{defTMDFF} 
D_{H/\mc{Q}} (z,\blp{q},\mu,\nu) = \sum_X \frac{1}{2N_c z} \mr{Tr} \langle 0 | \delta \left(\frac{p_+}{z} - \mc{P}_+ \right) \delta^{(D-2)} (\blp{q} -\bsp{\mP}) \nn \Psi_n^{\mQ} | H(p) X \rangle  \langle H(p) X | \bar{\Psi}_n^{\mQ} | 0 \rangle . 
\ee
Here the fragmenting process is described by $n$-collinear interactions, where $n^{\mu} = (1,\hat{\bl{n}})$ and $\n^{\mu}= (1,-\hat{\bl{n}})$ are the lightcone vectors normalized to $n\cdot \n =2$. 
$\Psi_n^{\mQ} = W_n^{\dagger} \xi_n^{\mQ}$ is the gauge invariant massive quark field accompanying the collinear Wilson line, and $H$ is the hadron containing the heavy quark. 
$\mc{P}_+ \equiv \n\cdot \mc{P}$ and $\bsp{\mP}$ are the derivative operators that return the large momentum component and the transverse momentum respectively. $N_c$ is a number of colors and $\mu~(\nu)$ is an ordinary (rapidity) renormalization scale. 

In Eq.~\eqref{defTMDFF}, $\qp$ is the transverse momentum of an initiating parton with respect to the hadron momentum $p$. If we consider the fragmentation in the parton frame with the transverse momentum of the initiating parton set to zero, the fragmentation can be described as the distribution of the hadron's transverse momentum $\pp$ with reference to the initiating parton's momentum. 
The transverse momenta between the hadron and the parton frames have the relation 
\be 
\label{relfra} 
\qp = -\frac{\pp}{z},
\ee
where $z=p_+/q_+$ is the energy fraction of the hadron over the initial parton. In this section we will consider the fragmenting process over the whole range of $z$, but $z$ will be treated as neither much less than nor too close to 1. 
If the initiating heavy quark's transverse momentum with respect to the final hadron's momentum is comparable with the heavy quark mass $m$, i.e., $q_{\perp} \equiv |\blp{q}| \sim m$, 
the $n$-collinear interactions scale as 
\be 
\label{nsca} 
p_n^{\mu} = (\n\cdot p_n, n\cdot p_n,\blpu{p}_n) = (p_n^+,p_n^-,\blpu{p}_n) \sim Q(1, m^2/Q^2,m/Q), 
\ee
where $Q$ is a typical hard scale taken to be much larger than $m$. 

\begin{figure}[t]
\begin{center}
\includegraphics[height=3.6cm]{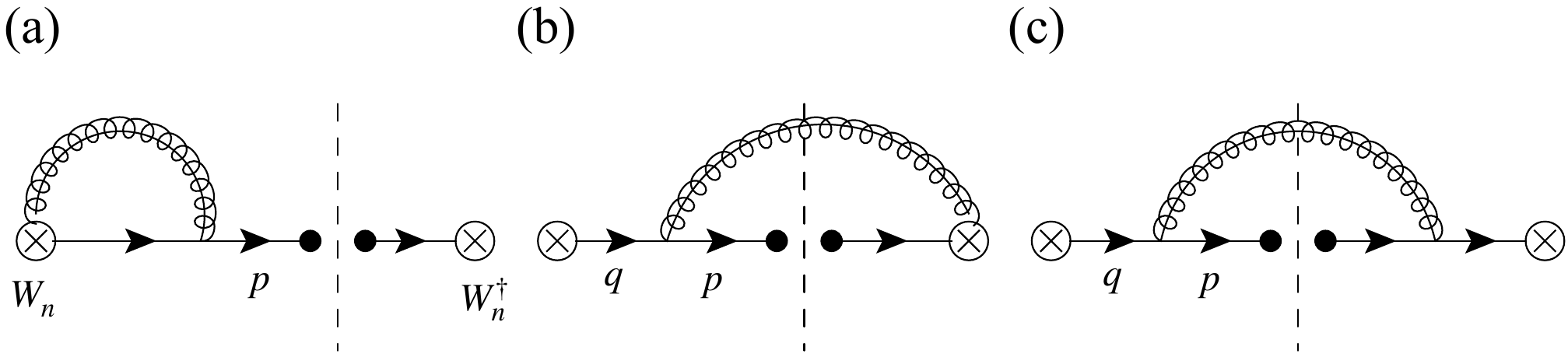}
\end{center}
\vspace{-0.8cm}
\caption{\label{fig1}
One-loop diagrams for calculation of the heavy quark TMDFF with $\blp{q} \sim m$. The vertical dashed lines are the unitary cuts. The self-energy diagrams of the heavy quark are omitted here. Diagrams (a) and (b) have mirror diagrams. 
}
\end{figure}

For the rest of this section, let us consider the one-loop calculation of the fragmentation function at parton level, i.e., $D_{\mQ/\mQ}$. 
%at hadron frame.
%with $q_{\perp} \sim m$. 
%(The one loop calculation of the function at parton frame will be shown in Appendix.~\ref{FFp}.) 
From this calculation, we will be able to extract the renormalization behavior of the fragmentation function with a heavy quark setting aside nonperturbative effects.   
At leading order (LO) in $\as$, the fragmentation function at the parton level is normalized as   
\be 
D_{\mQ/\mQ}^{(0)} (z,\blp{q}) = \delta(1-z) \delta^{(2)} (\blp{q}). 
\ee
At next-to-leading order (NLO) in $\as$, the one-loop diagrams are illustrated in Fig.~\ref{fig1}. 
To regularize the UV and the IR divergences in each diagram, we employ  on-shell dimensional regularization with $D=4-2\eps$. When we regularize the rapidity divergences in the heavy quark collinear sector, we use the conventional method~\cite{Chiu:2011qc,Chiu:2012ir} to modify the collinear Wilson line to\footnote{Then, following the prescription developed in Ref.~\cite{chay:2020jzn}, we will regularize the corresponding rapidity divergences in the soft sector.
}   
\be
\label{tilWil} 
W_n = \sum_{\mr{perm}} \exp \Bigl[-\frac{g}{\mP_+} \Bigl(\frac{\nu}{|\mP^g_+|} \Bigr)^{\eta} \n\cdot A_n \Bigr]. 
\ee

As discussed in Ref.~\cite{chay:2020jzn}, the rapidity divergences originate from the fact that the soft degrees of freedom cannot describe the large rapidity region. Hence, in the calculation of the collinear heavy quark sector, naive collinear contributions  do not yield the divergences. Instead, the rapidity divergences occur in the zero-bin contribution~\cite{Manohar:2006nz}, which needs to be subtracted in order to avoid  double counting  the soft contributions.  

The virtual contribution for Fig.~\ref{fig1}(a) has been computed in Ref.~\cite{chay:2020jzn}. Including the zero-bin subtraction, the result is 
\bea
\label{colMa} 
M_a &=& \frac{\as C_F}{4\pi} \Biggl[\frac{2}{\euv} + 2\ln \frac{\mu^2}{m^2} + \frac{1}{\eir^2} +\frac{1}{\eir} \ln\frac{\mu^2}{m^2} 
+\frac{1}{2} \ln^2 \frac{\mu^2}{m^2} + 4 + \frac{\pi^2}{12} \nnb \\
&&~~~~~~~~+\Bigl(\frac{2}{\eta} + 2 \ln\frac{\nu}{p_+} \Bigr) \Bigl(\frac{1}{\euv} -\frac{1}{\eir} \Bigr)\Biggr], 
\eea
where $p_+ \sim 2 E$ is the largest momentum component of the heavy quark in the final state. 
The rapidity scale that minimizes the large logarithm with $p_+$ is $\nu_c \sim p_+$. 
Including the mirror contribution of Fig.~\ref{fig1}(a) and combining with the self-energy contributions, 
\be 
Z_{\mQ}^{(1)} + R_{\mQ}^{(1)} = - \frac{\as C_F}{4\pi} \Bigl(\frac{1}{\euv} + \frac{2}{\eir} + 3\ln\frac{\mu^2}{m^2} +4\Bigr), 
\ee
the overall virtual contribution is
\bea 
\mc{M}_{V}(z,\blp{q}) &=& [2 M_a +Z_{\mQ}^{(1)} + R_{\mQ}^{(1)}] \delta(1-z) \delta^{(2)} (\blp{q}) \nnb \\
&=& \frac{\as C_F}{2\pi} \Biggl[\frac{3}{2\euv} +\frac{1}{\eir^2} +\frac{1}{\eir}\Bigl(-1+ \ln\frac{\mu^2}{m^2}\Bigr) +\frac{1}{2} \ln \frac{\mu^2}{m^2} +  
+\frac{1}{2} \ln^2 \frac{\mu^2}{m^2} + 2 + \frac{\pi^2}{12} \nnb \\
\label{colMV} 
&&~~~~~~~~+\Bigl(\frac{2}{\eta} + 2 \ln\frac{\nu}{p_+} \Bigr) \Bigl(\frac{1}{\euv} -\frac{1}{\eir} \Bigr)\Biggr]
\delta(1-z) \delta^{(2)} (\blp{q}). 
\eea 

The naive collinear contribution from the real emission in Fig.~\ref{fig1}(b) is 
\be
\label{colMb1} 
\tilde{M}_b (z,\blp{q}) = \frac{\as C_F}{2\pi^2} \frac{\mu^2 e^{\gamma_E}}{\Gamma(1-\eps)} \frac{z^{1-2\eps}}{1-z} \cdot \frac{(\blp{q}^2)^{-\eps}}{\blp{q}^2+\frac{(1-z)^2}{z^2}m^2}\ . 
\ee
Here, using the plus distribution, we re-express $z/(1-z)$  as 
\be
\frac{z}{1-z} 
%= \delta(1-z) \Biggl[\int^1_0 dz' \frac{z'}{1-z'} \Biggr] + \left(\frac{z}{1-z}\right)_+
= \delta(1-z) \Biggl[\int^1_0 dx \frac{1-x}{x} \Biggr] + \left(\frac{z}{1-z}\right)_+.
\ee
Then Eq.~\eqref{colMb1} can be rewritten as 
\bea
\label{colMb2} 
\tilde{M}_b (z,\blp{q}) &=& \frac{\as C_F}{2\pi^2} \frac{(\mu^2 e^{\gamma_E})^{\eps}}{\Gamma(1-\eps)} \Biggl[
\delta(1-z) \cdot \int^1_0 dx \frac{1-x}{x} \cdot \frac{1}{(\blp{q}^2)^{1+\eps}}  \\
&&\phantom{\frac{\as C_F}{2\pi^2} \frac{(\mu^2 e^{\gamma_E})^{\eps}}{\Gamma(1-\eps)} \Biggl[}
  +\left(\frac{z}{1-z}\right)_+ \cdot \frac{z^{-2\eps}(\blp{q}^2)^{-\eps}}{\blp{q}^2+\frac{(1-z)^2}{z^2}2m^2}\Biggr]\ . \nnb
\eea

To complete calculation, we need to subtract the zero-bin contribution that comes from the underlying soft interaction. 
Here the soft mode generally scales as 
\be 
\label{softsc} 
p_s^{\mu} = (p_s^+,p_s^-,\blpu{p}_s) \sim \Bigl(\frac{1}{\kappa} q_{\perp}, \kappa q_{\perp}, q_{\perp}\Bigr),~~q_{\perp} \sim m 
\ee
where the scaling of the boosting parameter $\kappa$ is given by 
\be 
\frac{m}{Q} \ll \kappa \lesssim 1. 
\ee
When $\kappa$ is in this range, the soft gluon radiations from the boosted $n$-collinear heavy quark eikonalize satisfying the approximation, $2p_n\cdot p_s \approx p_n^+ p_s^-$, giving rise to the soft Wilson line, 
\be 
S_n (x)= \mr{P} \exp \Bigl[ig \int^{\infty}_x ds n\cdot A_s (sn)\Bigr]. 
\ee 
With the scaling behavior of Eq.~\eqref{softsc} assigned, the zero-bin contribution for $\tilde{M}_b(z,\blp{q})$  is given by 
\be
\label{colMbzb} 
M_b^{\varnothing} (z,\blp{q}) = \frac{\as C_F}{2\pi^2} \frac{(\mu^2 e^{\gamma_E})^{\eps}}{\Gamma(1-\eps)} \left(\frac{\nu}{p_+}\right)^{\eta}
\left(\int^{\infty}_0 dx x^{-1-\eta}\right) \frac{1}{(\blp{q}^2)^{1+\eps}}\cdot \delta(1-z),  
\ee
where the upper limit of the integral for the gluon momentum fraction $x~(= k_+/p_+)$ has been set to infinity since
the soft gluon momentum $k_+$ in the zero-bin has no upper bound.  
The rapidity regulator, using Eq.~\eqref{tilWil}, will regulate the rapidity divergence as $k_+  \to \infty$. 

Subtracting Eq.~\eqref{colMbzb} from Eq.~\eqref{colMb2}, the soft divergence as $x\to 0$ cancel as follows: 
\begin{align}
\int^1_0 dx \frac{1-x}{x} - \left(\frac{\nu}{p_+}\right)^{\eta}\int^{\infty}_0 dx x^{-1-\eta}
&= \int^1_0 dx \Bigl[\frac{1-x}{x}-\frac{1}{x}\Bigr] - \left(\frac{\nu}{p_+}\right)^{\eta}\int^{\infty}_1 dx x^{-1-\eta} \nnb \\
\label{softcancel}
&= - 1 - \left(\frac{\nu}{p_+}\right)^{\eta} \frac{1}{\eta} \ .
\end{align}
Here $\eta~(\to +0)$ is a small positive number, hence its dependence can be suppressed in the integral region $x\in[0,1]$. 
So, after the subtraction, the complete contribution for Fig.~\ref{fig1}(b) is given as 
\bea 
M_b (z,\blp{q}) &=& \tilde{M}_b (z,\blp{q}) - M_b^{\varnothing} (z,\blp{q}) \nnb \\
\label{colMb3} 
&=& \frac{\as C_F}{2\pi^2} \frac{(\mu^2 e^{\gamma_E})^{\eps}}{\Gamma(1-\eps)} \Biggl[
-\delta(1-z) \Bigl(\frac{1}{\eta} + \ln \frac{\nu}{p_+} +1\Bigr) \frac{1}{(\blp{q}^2)^{1+\eps}}\\
&& \phantom{ \frac{\as C_F}{2\pi^2} \frac{(\mu^2 e^{\gamma_E})^{\eps}}{\Gamma(1-\eps)} \Biggl[}
+\left(\frac{z}{1-z}\right)_+ \cdot \frac{z^{-2\eps}(\blp{q}^2)^{-\eps}}{\blp{q}^2+\frac{(1-z)^2}{z^2}m^2}\Biggr]. \nnb
\eea 
Here the term $1/(\blp{q}^2)^{1+\eps}$ has a collinear IR divergence when $\blp{q}^2 \to 0$. In order to isolate the divergence we rewrite it as 
\bea 
\frac{1}{(\blp{q}^2)^{1+\eps}} &=& \delta (\blp{q}^2) \Biggl[\int^{\Lambda^2}_0  d\blp{l}^2 (\blp{l}^2)^{-1-\eps}\Biggr] 
+ \Biggl[\frac{1}{(\blp{q}^2)^{1+\eps}} \Biggr]_{\Lambda^2} \nnb \\
\label{ld1} 
&=& \delta (\blp{q}^2) \Bigl(-\frac{1}{\eir} + \ln \Lambda^2 \Bigr) + \Biggl[\frac{1}{\blp{q}^2} \Biggr]_{\Lambda^2} + \mc{O}(\eps),
\eea
where $[\cdots]_{\Lambda^2}$ is the so-called $\Lambda^2$-distribution, which is the dimensionful plus distribution, defined by 
\be
\label{ldd} 
[g(\qps)]_{\Lambda^2} = g(\qps) - \delta(\qps) \int^{\Lambda^2}_0 d\blp{l}^2 g(\blp{l}^2). 
\ee
Here $\delta (\qps) = \pi\delta^{(2)} (\blp{q})$, and $\Lambda^2$ is an arbitrary momentum squared scaling as $\sim \qps$. The overall calculation does not depend on any particular choice of $\Lambda^2$ as we will see.  
 
For the second term in the square bracket of Eq.~\eqref{colMb3}, we also employ the $\Lambda^2$-distribution by rewriting 
\be 
\label{ld2} 
\frac{(z^2\blp{q}^2/\mu^2)^{-\eps}}{\blp{q}^2+\frac{(1-z)^2}{z^2} m^2} = f\bigl(z,\lambda;\frac{\mu^2}{\Lambda^2}\bigr) \delta(\qps) + \Biggl[\frac{(z^2\blp{q}^2/\mu^2)^{-\eps}}{\blp{q}^2+\frac{(1-z)^2}{z^2} m^2}\Biggr]_{\Lambda^2}, 
\ee 
where $f(z,\lambda;\mu^2/\Lambda^2)$ is defined by the following integral 
\be 
\label{fzla} 
f\bigl(z,\lambda;\frac{\mu^2}{\Lambda^2}\bigr) = \mu^{2\eps} \int^{\Lambda^2}_0 d\blp{l}^2  
\frac{(z^2\blp{l}^2)^{-\eps}}{\blp{l}^2+\frac{(1-z)^2}{z^2}m^2} = \Bigl(\frac{\mu^2}{\Lambda^2} \Bigr)^{\eps} \int^{z^2}_0 dy \frac{y^{-\eps}}{y+(1-z)^2 \lambda}~, 
\ee
with $\lambda \equiv m^2/\Lambda^2$. $f(z,\lambda)$ becomes divergent as $z$ goes to 1. Thus, in order to extract the IR divergences fully, we rewrite the combination of $[z/(1-z)]_+$ and $f(z,\lambda)$ in Eq.~\eqref{colMb3} by  
\be
\label{frel} 
\left(\frac{z}{1-z}\right)_+ f(z,\lambda) =  \left(\frac{z f(z,\lambda)}{1-z}\right)_+ + \delta(1-z) \int^1_0 dz' \Bigl(\frac{z'}{1-z'}\Bigr)\Bigl[f(z',\lambda)-f(1,\lambda)\Bigr].  
\ee
Here the first term in the right side is finite $z \to 1$, 
and the integration in the second term is
\bea 
&&\frac{e^{\gamma_E}}{\Gamma(1-\eps)} 
\int^1_0 dz \Bigl(\frac{z}{1-z}\Bigr)\Bigl[f(z,\lambda)-f(1,\lambda)\Bigr] \nnb \\ 
\label{fintir} 
&&~~~
= -\frac{1}{2\eir^2} -\frac{1}{\eir} \Bigl(1+\frac{1}{2} \ln \frac{\mu^2}{m^2} \Bigr) 
- \ln\frac{\mu^2}{m^2} -\frac{1}{4} \ln^2\frac{\mu^2}{m^2} - \frac{\pi^2}{24} 
\\
&&~~~~~
-\frac{2}{\sqrt{\lambda}} \arctan{\sqrt{\lambda}} -\ln(1+\lambda)  -\frac{1}{2} \mr{Li}_2 (-\lambda) - F(\lambda),
\nnb 
\eea 
where $F(\lambda)$ has the following integral form, 
\be 
F(\lambda) =\int^1_0 dz \frac{z}{1-z} \int^1_{z^2} dy \frac{1}{y+(1-z)^2 \lambda}\ ,   
\ee
and $F(\lambda=1) = -\ln 2 + \pi^2/6$ and $F(0) = -2 + \pi^2/3$.

Finally, putting Eqs.~\eqref{ld1} and \eqref{ld2} into Eq.~\eqref{colMb3} and using the results in Eqs.~\eqref{frel} and \eqref{fintir}, we obtain the real contribution for Fig.~\ref{fig1}(b),  
\bea 
\label{colMb4} 
M_b (z,\blp{q}) &=&  \frac{\as C_F}{2\pi^2}
\Biggl\{ \delta(1-z) \Bigl(\frac{1}{\eta} + \ln \frac{\nu}{p_+} +1\Bigr)\Bigl[\delta(\qps)\Bigl(\frac{1}{\eir} + \ln \frac{\mu^2}{\Lambda^2} \Bigr) - \Bigl(\frac{1}{\qps}\Bigr)_{\Lambda^2} \Bigr]
 \\
&&-\delta(1-z)\delta(\qps) \Bigl[\frac{1}{2\eir^2} +\frac{1}{\eir} \Bigl(1+\frac{1}{2} \ln \frac{\mu^2}{m^2} \Bigr) 
+ \ln\frac{\mu^2}{m^2} +\frac{1}{4} \ln^2\frac{\mu^2}{m^2} + \frac{\pi^2}{24} \nnb \\
&&+\frac{2}{\sqrt{\lambda}} \arctan{\sqrt{\lambda}} +\ln(1+\lambda)  +\frac{1}{2} \mr{Li}_2 (-\lambda)+F(\lambda)\Bigr]
\nnb \\
&&+\delta(\qps) \Bigl(\frac{z}{1-z}\ln\frac{z^2+(1-z)^2 \lambda}{(1-z)^2 \lambda}\Bigr)_+ 
+\Bigl(\frac{z}{1-z}\Bigr)_+ \Bigl(\frac{z^2}{z^2 \qps+(1-z)^2 m^2}\Bigr)_{\Lambda^2} \Biggr\}.  \nnb
\eea 
We have extracted all the possible IR divergences as $\qps \to 0$ or $z\to 1$ and assigned them to the term with $\delta(1-z) \delta(\qps)$. The remaining terms with either the plus or the $\Lambda^2$-distributions are IR finite. 

The contribution for the diagram in Fig.~\ref{fig1}(c) is given by  
\bea 
\label{Mcf} 
M_c (z,\blp{q}) &=& \frac{\as C_F}{2\pi^2} \frac{(\mu^2 e^{\gamma_E})^{\eps}}{\Gamma(1-\eps)} (1-z) z^{-2\eps} (\qps)^{-\eps} \\
&&\times \Biggl[\frac{1-\eps}{\qps + \frac{(1-z)^2}{z^2} m^2} - \frac{2m^2}{z (\qps + \frac{(1-z)^2}{z^2} m^2)^2} \Biggr] 
\equiv M_{c1} (z,\blp{q})+M_{c2} (z,\blp{q}), \nnb 
\eea 
where $M_{c1}~(M_{c2})$ corresponds to the contribution from the first (second) term in the square bracket. 
These contributions do not need zero-bin subtractions since the corresponding contribution from the soft mode is power-suppressed. 

Due to the presence of $(1-z)$ in the numerator, $M_{c1}$ has no IR divergence. So, ignoring the $\eps$ dependence, we obtain 
\be
\label{Mc1} 
M_{c1} (z,\blp{q}) = \frac{\as C_F}{2\pi^2} (1-z)\Biggl[ \ln \frac{z^2+(1-z)^2\lambda}{(1-z)^2\lambda} \cdot \delta(\qps)
+ \Bigl(\frac{z^2}{z^2\qps+(1-z)^2 m^2}\Bigr)_{\Lambda^2} \Biggr].  
\ee
$M_{c2}$ has an IR divergence as $z\to 1$ and $\qps \to 0$ simultaneously. Using the plus and the $\Lambda^2$-distributions we can extract the IR divergence,  leading to 
\bea 
M_{c2} (z,\blp{q})&=& \frac{\as C_F}{2\pi^2} \Biggl\{\delta(\qps) \Biggl[ \delta(1-z) \Bigl(\frac{1}{\eir} + \ln\frac{\mu^2}{m^2} + \frac{2}{\sqrt{\lambda}} \arctan{\sqrt{\lambda}} + \ln(1+\lambda) + G(\lambda) \Bigr)  \nnb \\
\label{Mc2} 
&&\hspace{-1cm} - \Bigl(\frac{2z^3}{(1-z)(z^2+(1-z)^2\lambda}\Bigr)_+ \Biggr] -2z^3(1-z) \Bigl(\frac{m^2}{(z^2\qps + (1-z)^2 m^2)^2} \Bigr)_{\Lambda^2} \Biggr\}. 
\eea 
Here $G(\lambda)$ has a form of the integral, 
\be 
G(\lambda) = \int^1_0 dz \int^1_{z^2} dy \frac{2z(1-z) \lambda}{(y+(1-z)^2\lambda)^2}\ , 
\ee
where $G(0) = 0$ and $G(1) = 1-\ln2$.

Finally, combining the results of Eqs.~\eqref{colMV}, \eqref{colMb4}, \eqref{Mc1}, and \eqref{Mc2}, we obtain the bare one-loop result for the heavy quark TMDFF,  
\bea 
D_{\mQ/\mQ}^{(1)} (z,\blp{q}) &=& \mc{M}_V(z,\blp{q}) + 2M_b (z,\blp{q}) + M_c(z,\blp{q}) \nnb \\
\label{DQQ1}
&=&\frac{\as C_F}{2\pi^2} \Biggl\{ \delta(1-z) \delta(\qps) 
\Biggl[\Bigl(\frac{2}{\eta} + 2\ln\frac{\nu}{p_+} +\frac{3}{2} \Bigr) \Bigl(\frac{1}{\euv} + \ln\frac{\mu^2}{\Lambda^2} \Bigr)+2 \\ 
&& -\ln(1+\lambda) - \frac{2}{\sqrt{\lambda}} \arctan{\sqrt{\lambda}} - \mr{Li}_2 (-\lambda) -F(\lambda) + G(\lambda) \Biggr] 
-\delta(\qps) \Biggl[\frac{P_{qq}(z)}{C_F} \ln\lambda \nnb \\
&&+\left(\frac{2z}{1-z} \Bigl(\ln\frac{z^2+(1-z)^2 \lambda}{(1-z)^2} - \frac{z^2}{z^2+(1-z)^2 \lambda}\Bigr)\right)_+
+ (1-z) \ln\frac{z^2+(1-z)^2 \lambda}{(1-z)^2}\Biggr] \nnb \\ 
&& -\Bigl(\frac{2}{\eta} + 2\ln\frac{\nu}{p_+} +\frac{3}{2} \Bigr) \delta(1-z)\left(\frac{1}{\qps}\right)_{\Lambda^2} 
+ \frac{P_{qq}(z)}{C_F} \left(\frac{z^2}{z^2\qps+(1-z)^2 m^2}\right)_{\Lambda^2} \nnb \\
&&
-2z^3(1-z) \left(\frac{m^2}{(z^2\qps + (1-z)^2 m^2)^2} \right)_{\Lambda^2} \Biggr\}\ . \nnb 
\eea 
Here $P_{qq}$ is quark-to-quark DokshitzerGribov-Lipatov-Altarelli-Parisi (DGLAP) kernel, 
\be 
P_{qq}(z) = C_F \left( \frac{1+z^2}{1-z} \right)_+.  
\ee 
As shown in Eq.~\eqref{DQQ1}, the heavy quark TMDFF is IR finite since the IR divergences from the real emission contributions, $2M_b+M_c$, are cancelled by the virtual contribution $\mc{M}_V$.  
The contributions proportional to $P_{qq}$ involve the logarithm of the heavy quark mass, i.e., $\ln \lambda = \ln m^2/\Lambda^2$. If we consider the massless limit of the heavy quark, they become collinear-divergent.   

Note that UV divergence of the heavy quark TMDFF genuinely comes from the virtual contribution, which makes sense since the real emission contributions with a finite $q_{\perp}$ cannot produce a UV divergence. Comparing to the light quark calculation, we expect the same UV divergence since the inclusion of the quark mass cannot change the UV behavior. The presence of the fermion mass does change the IR behavior and makes it possible to compute the HQ TMDFF perturbatively. 
Finally, the rapidity divergence for the heavy quark TMDFF is  the same as the light-quark case, since the rapidity divergence comes from the zero-bin contributions in the soft sector, %scaling as Eq.~\eqref{softsc}, 
which is common for both. 

When we consider a generic $N$-jet process, it is useful to introduce multiple rapidity scales $\nu_{i}~(i=1,\cdots,N)$ corresponding to the separated $N$ collinear directions~\cite{chay:2020jzn}. In this case, the anomalous dimensions for TMDFFs satisfy the following renormalization group (RG) equations:
\bea 
\frac{d}{d\ln\mu} D_{f/f} (z,\blp{q},\mu,\nu_i) &=&  \gamma_{f}^{\mu} (\mu,\nu_i) D_{f/f} (z,\blp{q},\mu,\nu_i), \\
\frac{d}{d\ln\nu_i} D_{f/f} (z,\blp{q},\mu,\nu_i) &=& \int d^2 \blp{l} 
\gamma_{f}^{\nu} (\blp{l};\mu,\nu_i) D_{f/f} (z,\blp{q}-\blp{l},\mu,\nu_i), \nnb 
\eea 
with
\bea 
\label{gfmu}
\gamma_{f}^{\mu} (\mu,\nu_i) &=& \frac{\as}{\pi} \Bigl( \bl{T}_f^2 \cdot 2\ln\frac{\nu_i}{p_i^+} + \frac{\hat{\gamma}_f}{2} \Bigr) + \mc{O}(\as^2),   \\
\label{gfnu}
\gamma_{f}^{\nu} (\blp{q},\mu,\nu_i) &=& \frac{\as}{\pi^2} \bl{T}_f^2 \Biggl[\ln\frac{\mu^2}{\Lambda^2} \cdot \delta(\qps) - \Bigl(\frac{1}{\qps}\Bigr)_{\Lambda^2}\Biggr]+ \mc{O}(\as^2). 
\eea 
Here $\bl{T}_f^2 = \bl{T}_f^a \cdot \bl{T}_f^a$ becomes $C_F$ for $f = q$~(quark) and $C_A$ for $f=g$~(gluon). 
$\hat{\gamma}_q = 3C_F$ and $\hat{\gamma}_g = \beta_0$, where $\beta_0$ is the leading coefficient of QCD beta function. 
In Eq.~\eqref{gfnu} we employed $\Lambda^2$-distribution introduced in Eq.~\eqref{ldd} and the net result should be independent of $\Lambda^2$. 

In the impact parameter $(\bl{b})$ space, the heavy quark TMDFF can be expressed 
through the Fourier transform, 
\be 
\tilde{D}_{H/\mQ} (z,\bl{b},\mu,\nu) = \int d^2\blp{q} e^{i\bl{b} \cdot \blp{q}} D_{H/\mQ} (z,\blp{q},\mu,\nu). 
\ee
In  $\bb$-space, the renormalized result at NLO is 
\bea 
\tilde{D}_{\mQ/\mQ} (z,\bb,\mu,\nu) &=& 1+ \frac{\as C_F}{2\pi}  \Biggl\{ \delta(1-z) \Bigl[\Bigl(2\ln\frac{\nu}{p_+}+\frac{3}{2} \Bigr) \ln \bar{b}^2 \mu^2 + \frac{1}{2} \ln \bar{b}^2 m^2 \Bigr] \nnb \\
\label{DQQ1b} 
&&+\Bigl(\frac{2z}{1-z}\Bigr)_+ \Bigl[2K_0 \Bigl(\frac{1-z}{z}mb\Bigr) +2\ln(1-z) -1\Bigr] \\
&&+2(1-z)K_0 \Bigl(\frac{1-z}{z}mb\Bigr)-\Bigl(\frac{4z}{1-z} \ln(1-z)\Bigr)_+  \nnb \\
&&-2(1-z) \Bigl[\frac{bm}{1-z} K_1 \Bigl(\frac{1-z}{z}mb\Bigr)-\frac{z}{(1-z)^2} \Bigr] \Biggr\}, \nnb
\eea 
where $b^2=\bb^2$ and $\bar{b}\equiv be^{\gamma_E}/2$. 
$K_{n=0,1}$ are the modified Bessel functions of the second kind. 
As $z$ goes to 1, the following combinations with the Bessel functions remain nonsingular: 
\bea 
\label{relK0} 
K_0 \Bigl(\frac{1-z}{z}mb\Bigr) +\ln(1-z) &=&   - \ln m \bar{b}   + \mc{O}(1-z), \\
\label{relK1} 
\frac{bm}{1-z} K_1 \Bigl(\frac{1-z}{z}mb\Bigr)-\frac{z}{(1-z)^2} &=& \frac{m^2b^2}{4} \Bigl(-1+2\ln (1-z) m\bar{b}
\Bigr)+ \mc{O}(1-z). 
\eea 
Finally, the leading anomalous dimension for $\tilde{D}_{\mQ/\mQ}$ satisfying the RG equation, 
$\frac{d}{d\ln\nu}\tilde{D}_{\mQ/\mQ} = \tilde{\gamma}_{\mQ}^{\nu}\cdot \tilde{D}_{\mQ/\mQ}$ in $\bb$-space is given by 
\be
\label{tgQnu}
\tilde{\gamma}_{\mQ}^{\nu} (\bb;\mu,\nu) = \frac{\as C_F}{\pi} \ln \bar{b}^2 \mu^2 \ . 
\ee

\section{The heavy quark TMD fragmentation function for $q_{\perp} \ll m$} 
\label{smallq} 

In this section, we  consider the region of parameter space where the transverse momentum $q_{\perp}$ is much smaller than the heavy quark mass $m$, so the fluctuations to describe $q_{\perp}$ should be much softer than the collinear interaction scaling shown in Eq.~\eqref{nsca}.  Therefore, the heavy quark can be considered to be boosted, and we can integrate out the collinear interactions.
In this boosted heavy quark system, with the collinear interaction being integrated out, the remaining fluctuations are described by the residual interaction, where the momentum scales as   
\be 
\label{resim} 
k^{\mu} = (k_+,k_-,\blp{k}) \sim \veps Q (1,m^2/Q^2,m/Q).
\ee 
Here the small parameter $\veps$ has the size $\veps \sim q_{\perp}/m \ll 1$.  

This residual interaction can be systematically analyzed in the boosted heavy quark effective theory (bHQET), which can be directly obtained from the massive version of SCET ($\mr{SCET_M}$)~\cite{Leibovich:2003jd,Rothstein:2003wh,Chay:2005ck}. At leading power in the heavy quark limit, the bHQET Lagrangian is given by~\cite{Kim:2020dgu,Dai:2021mxb,Beneke:2023nmj} 
\be 
\label{LbHQET} 
\mc{L}^{(0)}_{\mr{bHQET}} = \bar{h}_n v\cdot iD \nn h_n, 
\ee
where the boosted heavy quark spinor satisfies the same projection as the spinor in SCET, 
\be 
\fms{n} h_n = 0,~~\frac{\fms{n}\fms{\n}}{4} h_n = h_n.  
\ee 
The velocity in Eq.~\eqref{LbHQET} scales as $v^{\mu} = (v_+,v_-,\blp{v}) \sim (Q/m,m/Q,1)$ and is normalized to $v^2 =1$.   

Therefore, when $q_{\perp} \ll m$, the HQ TMDFF in Eq.~\eqref{defTMDFF} can be matched onto bHQET and can be factorized as 
\be 
\label{HQFFF} 
D_{H/\mc{Q}} (z,\blp{q}\ll m;\mu,\nu) = C_{\mQ} (m,\mu) S_{H/\mQ} (z,\blp{q},\mu,\nu). 
\ee
Here $C_{\mQ}$ is the matching coefficient onto bHQET obtained from integrating out the virtual collinear interaction, which at NLO is~\cite{Fleming:2007xt,Neubert:2007je,Fickinger:2016rfd} 
\be 
\label{CQnlo} 
C_{\mQ} (m,\mu) = 1+\frac{\as C_F}{4\pi} \Bigl(\ln\frac{\mu^2}{m^2} + \ln^2\frac{\mu^2}{m^2} + 4 + \frac{\pi^2}{6} \Bigr). 
\ee

$S_{H/\mQ}$ is the HQTMD shape function to be described within bHQET, which can be obtained through the direct matching from Eq.~\eqref{defTMDFF}, 
\bea 
S_{H/\mQ} (z,\blp{q},\mu,\nu) &=& \sum_{X_r} \frac{1}{2N_c} \mr{Tr} \frac{v_+}{2} \langle 0 | \delta\bigl( \frac{p_+}{z} - mv_+ - i\partial_+ \bigr) \delta^{(2)} (\blp{q} -\bsp{\mP}) Y_{\n}^{r\dagger} h_n | H(p) X_r \rangle  \nnb \\
\label{SHdef}
&&~~~\times \langle H(p) X_r | \bar{h}_n Y_{\n}^r \nn  |0\rangle, 
\eea 
where $X_r$ denotes the final states of the residual modes,
%scaling as Eq.~\eqref{resim}. 
and $Y_{\n}^r$ is the Wilson line of the residual gluons, which has been matched from $W_n$ with collinear gluons integrated out. 
Here we set the momentum of the final hadron as $p^{\mu} = m_H v^{\mu}$ with $\blp{v} =0$, where $m_H$ is the hadron mass,  
and the momentum of the initial mother heavy quark is given by $q^{\mu} = m v^{\mu} + k^{\mu}$. 
Correspondently, the scaling of the transverse momentum is given by $\blp{q} = \blp{k} \sim \veps m \ll m$.
$i\partial_+$ in the argument of the delta function in Eq.~\eqref{SHdef} takes the 
residual momentum of the initial heavy parton, $k_+$ and scales as $k_+ \sim \veps Q \ll Q \sim q_+ (= mv_+)$. 
So the argument of the delta function  holds when $z$ is close to 1, and it can be written as 
\be
\frac{p_+}{z} - m v_+ - k_+ = \frac{(1-z)mv_+}{z} + \frac{\bar{\Lambda} v_+}{z} - k_+ \sim (1-z) mv_+ + \bar{\Lambda} v_+ - k_+, 
\ee
where $\bar{\Lambda} = m_H - m \sim \mc{O}(\Lambda_{\mr{QCD}})$. Hence this shape function for $\qp \ll m$ in Eq.~\eqref{SHdef} has support in a large $z$ region. 
At the parton level $(H=\mQ)$, the argument of the delta function in the shape function becomes $(1-z) m v_+ - i\partial_+$, and, 
at LO in $\as$, the shape function is normalized to 
\be 
S_{\mQ/\mQ}^{(0)} (z,\blp{q}) = \delta(1-z) \delta^{(2)} (\blp{q}).  
\ee 
In obtaining this, we used the following spin sum rule for the boosted heavy quark field~\cite{Dai:2021mxb},  
\be 
\sum_s h_n | \mQ_s \rangle \langle \mQ_s | \bar{h}_n = m \fms{n}. 
\ee

%The dominant velocity component $v_+$ is here approximately given by $p_+/m$.\footnote{
%Here, we set the momenta of the initial mother parton and the final hadron as $q^{\mu}= mv^{\mu} + k^{\mu}$ and $p^{\mu}= Mv^{\mu}$ respectively, where $M$ is the hadron mass.  In the hadron frame with $\pp =0$, the transverse component of the velocity, $\blp{v}$, is given by zero. Correspondently, the scaling of the transverse momentum $\qp$ is given as $\blp{q} = \blp{k} \sim \veps m$. }
%similarly with  Eq.~\eqref{defTMDFF}, we have set the frame for the initial heavy mother parton to have zero transverse momentum. Hence, transverse component of the velocity $\blp{v}$ is given by zero.  If we write the full momentum of the observed hadron as $p^{\mu} = m v^{\mu} + k^{\mu}$, we easily see the scaling of the transverse momentum as $\blp{p} = \blp{k} \sim \veps m$. }

\begin{figure}[t]
\begin{center}
\includegraphics[height=3cm]{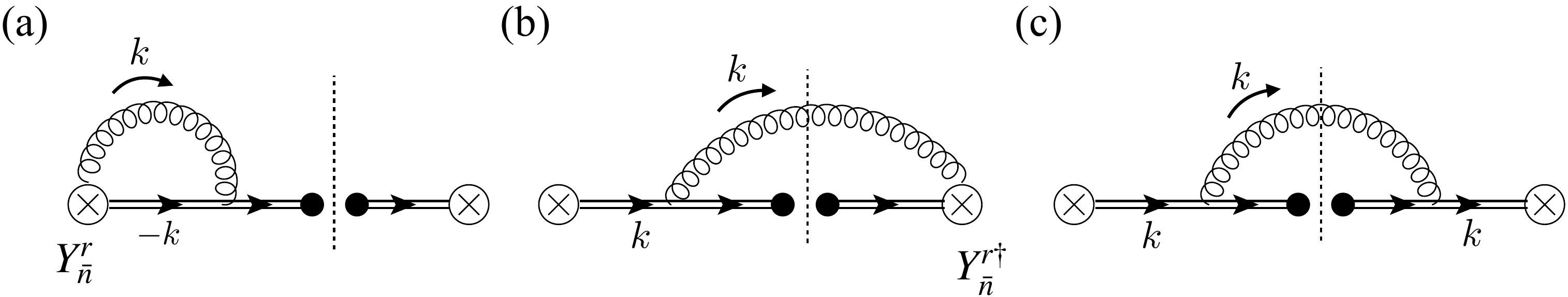}
\end{center}
\vspace{-0.8cm}
\caption{\label{fig2}
One-loop diagrams for calculating of the heavy quark shape function in bHQET for the $\mQ \to \mQ$ process. The mirror diagrams for diagrams (a,b) and the self-energy diagram for the heavy quark field are not shown here. 
The momentum for the final state is $p^{\mu} = mv^{\mu}$ with $\blp{v} = 0$, while the initial state heavy quark in
the real emission diagrams (b, c) has 
 momentum $q^{\mu} = mv^{\mu} + k^{\mu}$ with 
$\blp{q} = \blp{k}$.  }
\end{figure}

Let us consider the one-loop calculation of  the shape function at the parton level. 
The relevant Feynman diagrams are illustrated in Fig.~\ref{fig2}. 
The virtual contribution corresponding to Fig.~\ref{fig2}(a) is~\cite{chay:2020jzn} 
\bea 
M_a^{r} &=& \frac{\as C_F}{2\pi}  \Biggl[-\frac{1}{2} \Bigl(\frac{1}{\euv^2}-\frac{1}{\eir^2} \Bigr) 
-\frac{1}{2} \ln \frac{\mu^2}{m^2} \Bigl(\frac{1}{\euv} - \frac{1}{\eir} \Bigr) \nnb \\
\label{MabH} 
&&\phantom{\frac{\as C_F}{2\pi}  \Biggl[}
+\Bigl(\frac{1}{\eta} + \ln \frac{\nu}{p_+} \Bigr) \Bigl(\frac{1}{\euv} - \frac{1}{\eir} \Bigr) \Biggr]. 
\eea 
Like the collinear virtual contribution shown in Eq.~\eqref{colMa}, the bHQET result has also a rapidity divergence, which comes from the zero-bin contribution to be subtracted in the bHQET calculation. 
Note that the residual interaction scaling as Eq.~\eqref{resim} has almost the same rapidity as the collinear interaction although the residual mode has smaller energy.    
Hence, similar to the calculation of the collinear interaction, when we consider the bHQET calculation for the large rapidity region, we need to subtract the contribution of the small rapidity, 
i.e., the soft contribution. Here the soft interaction is supposed to scale as Eq.~\eqref{softsc}, but 
the offshellness is much smaller than $m^2$ since $q_{\perp} \ll m$. 
%of which interaction in general scales as\footnote{
%We call this degree of freedom the `ultrasoft (usoft)' mode, which is responsible for the small rapidity region when we consider the boosted residual momentum with the large rapidity. Hence its offshellness is given to be much smaller than the soft mode, i.e., $p_{us}^2 \ll p_s^2 \sim m^2$. } 
%\be 
%\label{scus} 
%(p_{us}^+, p_{us}^-, \blpu{p}_{us}) \sim \Bigl(\frac{1}{\kappa} q_{\perp}, \kappa q_{\perp}, q_{\perp}\Bigr),~~q_{\perp} \ll m,~~\frac{m}{Q} \ll \kappa \lesssim 1. 
%\ee 

For Fig.~\ref{fig2}(b), the naive contribution before the zero-bin subtraction is 
\be
\label{rMb1} 
\tilde{M}_b^r (z,\blp{q}) = \frac{\as C_F}{2\pi^2} \frac{\mu^2 e^{\gamma_E}}{\Gamma(1-\eps)} \frac{1}{1-z} \cdot \frac{(\blp{q}^2)^{-\eps}}{\blp{q}^2+(1-z)^2m^2}. 
\ee
When compared with Eq.~\eqref{colMb1}, Eq.~\eqref{rMb1} can be understood to be in the large $z$ region since the residual gluon emission has small energy, $k_+ \sim \veps Q$. 
This means that $(1-z)m$ and $\blp{q}$ can be power-counted as the same order, i.e., $(1-z) m \sim q_{\perp} \sim \veps m$. As in Eq.~\eqref{colMb1}, Eq.~\eqref{rMb1} becomes IR-divergent as $z \to 1$. To isolate the divergence, we employ the plus distribution for $1/(1-z)$, 
\be
\label{rMb2} 
\tilde{M}_b^r (z,\blp{q}) = \frac{\as C_F}{2\pi^2} \frac{(\mu^2 e^{\gamma_E})^{\eps}}{\Gamma(1-\eps)} \Biggl[
\delta(1-z) \left(\int^1_0 \frac{dx}{x} \right) \frac{1}{(\blp{q}^2)^{1+\eps}}  
+\frac{1}{(1-z)_+} \frac{(\blp{q}^2)^{-\eps}}{\blp{q}^2+(1-z)^2m^2}\Biggr]. 
\ee

The zero-bin contribution for Fig.~\ref{fig2}(b)  is given by 
\be
\label{rMbzb} 
M_b^{r,\varnothing} (z,\blp{q}) = \frac{\as C_F}{2\pi^2} \frac{(\mu^2 e^{\gamma_E})^{\eps}}{\Gamma(1-\eps)} \left(\frac{\nu}{p_+}\right)^{\eta}
\int^{\infty}_0 dx x^{-1-\eta} \frac{1}{(\blp{q}^2)^{1+\eps}}\cdot \delta(1-z),  
\ee
where the plus component of the soft gluon momentum is given by $p_{s}^+=xp_+$ and is assumed to be much smaller than the residual momentum, $k_+ \sim \veps Q$. Hence this zero-bin contribution only contributes to the part proportional to $\delta (1-z)$. In the integral the momentum fraction $x$  can reach infinity, so this contribution will involve a rapidity divergence.  

Subtracting Eq.~\eqref{rMbzb} from Eq.~\eqref{rMb2}, we remove the soft divergence as $x\to 0$ in a similar way as was shown in Eq.~\eqref{softcancel}. 
Then the complete contribution for Fig.~\ref{fig2}(b) becomes 
\bea 
M_b^{r} (z,\blp{q}) &=& \tilde{M}_b^r - M_b^{r,\varnothing} \nnb \\ 
&=& \frac{\as C_F}{2\pi^2}  \Biggl\{
\delta(1-z) \Bigl(\frac{1}{\eta} + \ln \frac{\nu}{p_+}\Bigr) 
\Bigl[\delta(\qps)\Bigl(\frac{1}{\eir} + \ln \frac{\mu^2}{\Lambda^2} \Bigr) - \Bigl(\frac{1}{\qps}\Bigr)_{\Lambda^2} \Bigr] 
\nnb \\
\label{rMb3}
&&\phantom{\frac{\as C_F}{2\pi^2}  \Biggl\{}
+\frac{(\mu^2 e^{\gamma_E})^{\eps}}{\Gamma(1-\eps)}\cdot \frac{1}{(1-z)_+} \cdot \frac{(\blp{q}^2)^{-\eps}}{\blp{q}^2+(1-z)^2m^2}\Biggr\},
\eea 
where we applied Eq.~\eqref{ld1}, using the $\Lambda^2$-distribution to extract IR divergence as $\qps \to 0$.  
For the second term in the curly bracket, we can also use the $\Lambda^2$-distribution in the form
\be 
\label{ld3} 
\frac{(\blp{q}^2/\mu^2)^{-\eps}}{\blp{q}^2+(1-z)^2m^2} = h\bigl(z,\lambda;\frac{\mu^2}{\Lambda^2}\bigr) \delta(\qps) + \Biggl[\frac{(\blp{q}^2/\mu^2)^{-\eps}}{\blp{q}^2+(1-z)^2m^2}\Biggr]_{\Lambda^2}. 
\ee 
Here $h(z,\lambda;\mu^2/\Lambda^2)$ is expressed as the following integral 
\be 
\label{hzla} 
h\bigl(z,\lambda;\frac{\mu^2}{\Lambda^2}\bigr) = \mu^{2\eps} \int^{\Lambda^2}_0 d\blp{l}^2  \frac{(\blp{l}^2)^{-\eps}}{\blp{l}^2+(1-z)^2m^2} = \Bigl(\frac{\mu^2}{\Lambda^2} \Bigr)^{\eps} \int^1_0 dy \frac{y^{-\eps}}{y+(1-z)^2 \lambda}~, 
\ee
where $\lambda = m^2/\Lambda^2$. Note that $h(z,\lambda)$ becomes divergent as $z$ goes to 1. Thus, in order to extract the IR divergences, 
we rewrite the combination of $1/(1-z)_+$ and $h(z,\lambda)$ as  
\be
\label{hrel2} 
\frac{1}{(1-z)_+} h(z,\lambda) =  \left(\frac{h(z,\lambda)}{1-z}\right)_+ + \delta(1-z) \int^1_0 \frac{dz'}{1-z'}\Bigl[h(z',\lambda)-h(1,\lambda)\Bigr].  
\ee
The integral in the second term produces IR divergences,
\bea 
\label{hintir2} 
&&\frac{e^{\gamma_E}}{\Gamma(1-\eps)} 
\int^1_0 \frac{dz}{1-z}\Bigl[h(z,\lambda)-h(1,\lambda)\Bigr] \\ &&
= -\frac{1}{2\eir^2} -\frac{1}{2\eir} \ln \frac{\mu^2}{m^2} 
-\frac{1}{4} \ln^2\frac{\mu^2}{m^2} - \frac{\pi^2}{24}  -\frac{1}{2} \mr{Li}_2 (-\lambda).
\nnb 
\eea 
Finally, combining the above, $M_b^r$ can be written as 
\bea 
\label{rMb4} 
M_b^r (z,\blp{q}) &=&  \frac{\as C_F}{2\pi^2}
\Biggl\{ \delta(1-z) \Bigl(\frac{1}{\eta} + \ln \frac{\nu}{p_+} \Bigr)\Bigl[\delta(\qps)\Bigl(\frac{1}{\eir} + \ln \frac{\mu^2}{\Lambda^2} \Bigr) - \Bigl(\frac{1}{\qps}\Bigr)_{\Lambda^2} \Bigr]
 \\
&&-\delta(1-z)\delta(\qps) \Bigl[\frac{1}{2\eir^2} +\frac{1}{2\eir} \ln \frac{\mu^2}{m^2} 
 +\frac{1}{4} \ln^2\frac{\mu^2}{m^2} + \frac{\pi^2}{24} +\frac{1}{2} \mr{Li}_2 (-\lambda)\Bigr] \nnb \\
&&+\delta(\qps) \Bigl(\frac{1}{1-z}\ln\frac{1+(1-z)^2 \lambda}{(1-z)^2 \lambda}\Bigr)_+ 
+\Bigl(\frac{1}{1-z}\Bigr)_+ \Bigl(\frac{1}{\qps+(1-z)^2 m^2}\Bigr)_{\Lambda^2} \Biggr\}.\nnb 
\eea 

The contribution for Fig.~\ref{fig2}(c) is given by 
\be
\label{rMc1} 
M_c^r (z,\blp{q}) = -\frac{\as C_F}{\pi^2} \frac{(\mu^2 e^{\gamma_E})^{\eps}}{\Gamma(1-\eps)} 
\frac{(1-z)  m^2 (\qps)^{-\eps}}{(\qps + (1-z)^2 m^2)^2}. 
\ee
The zero-bin contribution to this term is power-suppressed and can be ignored. 
Eq.~\eqref{rMc1} becomes IR-divergent when $z\to 1$ and $\qps \to 0$ simultaneously. So employing the plus and $\Lambda^2$-distributions we extract IR divergence and obtain 
\bea 
\label{rMc2} 
M_{c}^r (z,\blp{q})&=& \frac{\as C_F}{2\pi^2} \Biggl\{\delta(\qps) \Biggl[ \delta(1-z) \Bigl(\frac{1}{\eir} + \ln\frac{\mu^2}{m^2} + \ln(1+\lambda) \Bigr)  \\
&&~~~~~~~ - \Bigl(\frac{2}{(1-z)(1+(1-z)^2\lambda}\Bigr)_+ \Biggr] -2(1-z) \Bigl(\frac{m^2}{(\qps + (1-z)^2 m^2)^2} \Bigr)_{\Lambda^2} \Biggr\}. \nnb 
\eea 

Finally, together with the self-energy contribution of $h_n$, 
\be 
Z_h^{(1)} + R_h^{(1)} = \frac{\as C_F}{2\pi} \Bigl(\frac{1}{\euv} - \frac{1}{\eir} \Bigr),  
\ee 
we obtain the complete one loop correction to $S_{\mc{Q}/\mQ}$,
\bea 
S_{\mQ/\mQ}^{(1)} (z,\blp{q}) &=& \Bigl[2 M_a^r + Z_h ^{(1)} + R_h^{(1)} \Bigr] \delta(1-z) \delta(\qps) 
+2M_b^r (z,\blp{q}) +M_c^r (z,\blp{q}) \nnb \\
\label{SQnlo} 
&=& \frac{\as C_F}{2\pi^2} \Biggl\{ \delta(1-z) \Bigl(\frac{2}{\eta} + 2\ln \frac{\nu}{p_+}\Bigr) 
\Bigl[\delta(\qps)\Bigl(\frac{1}{\euv} + \ln \frac{\mu^2}{\Lambda^2} \Bigr) - \Bigl(\frac{1}{\qps}\Bigr)_{\Lambda^2} \Bigr] \\
&&
+\delta(1-z) \delta(\qps) \Bigl[-\frac{1}{\euv^2} +\frac{1}{\euv} \Bigl(1-\ln\frac{\mu^2}{m^2} \Bigr)+\ln\frac{\mu^2}{m^2}- \frac{1}{2}\ln^2\frac{\mu^2}{m^2} -\frac{\pi^2}{12}\nnb \\
&&+\ln(1+\lambda) -\mr{Li}_2 (-\lambda)  \Bigr] +\delta(\qps) \Bigl[\frac{2}{1-z} \Bigl(\ln\frac{1+(1-z)^2 \lambda}{(1-z)^2\lambda} - \frac{1}{1+(1-z)^2\lambda}\Bigr) \Bigr]_+ \nnb \\
&&+\frac{2}{(1-z)_+} \Bigl(\frac{1}{\qps+(1-z)^2m^2}\Bigr)_{\Lambda^2} 
-2(1-z) \Bigl(\frac{m^2}{(\qps +(1-z)^2m^2)^2} \Bigr)_{\Lambda^2}\Biggr\}. \nnb 
\eea
Here, as we expect, we see that IR divergences exactly cancel. The remaining UV divergences arise entirely from the virtual contributions ($2 M_a^r + Z_h ^{(1)}$). Also note the rapidity divergence is the same as the one for the TMDFF with $\qp \sim m$ as shown in Eq.~\eqref{DQQ1}.

In  $\bl{b}$-space, the renormalized HQTMD shape function at NLO is given by 
\bea 
\tilde{S}_{\mQ/\mQ} (z,\bl{b};\mu,\nu) &=& \int d^2 \blp{q} e^{i\bl{b}\cdot \qp} S_{\mQ/\mQ} (z,\qp;\mu,\nu)  \nnb \\
&=& \delta(1-z)+ \frac{\as C_F}{2\pi}  \Biggl\{ \delta(1-z) \Bigl(2\ln\frac{\nu}{p_+} \ln \bar{b}^2 \mu^2 + \ln\frac{\mu^2}{m^2} 
- \frac{1}{2} \ln^2 \frac{\mu^2}{m^2} -\frac{\pi^2}{12} \Bigr)  \nnb \\ 
&&-\Bigl[\frac{2}{1-z} (1+2\ln(1-z))\Bigr]_+ 
+ \frac{4}{(1-z)_+} \Bigl[K_0 ((1-z)mb) +\ln(1-z)\Bigr] \nnb \\
\label{SQb1}
&&-2(1-z) \Bigl[\frac{bm}{1-z} K_1 ((1-z)mb) -\frac{1}{(1-z)^2} \Bigr]\Biggr\}. 
\eea 
Here $b$ is power-counted as $b \sim 1/q_{\perp} \sim 1/(\veps m)$, hence the combination $(1-z) mb$ is  of $\mc{O} (1)$. 
%So the last term in Eq.~\eqref{SQb1} is suppressed by $(1-z) \sim \veps$ and can be ignored. 
The leading anomalous dimensions from the RG equations, 
\be
\frac{d}{d\ln s}\tilde{S}_{\mQ/\mQ}(z,\bb;s) = \tilde{\gamma}^s_{r} \cdot \tilde{S}_{\mQ/\mQ}(z,\bb;s),~~s=\mu,\nu,
\ee
are given by 
\begin{align}
\label{grmu}
\tilde{\gamma}_r^{\mu} (\mu,\nu) &= \frac{\as C_F}{\pi}\Bigl(2 \ln \frac{m\nu}{p_+ \mu} +1\Bigr), \\
\label{grnu}
\tilde{\gamma}_r^{\nu} (\bb;\mu,\nu) &= \frac{\as C_F}{\pi} \ln \bar{b}^2 \mu^2.  
\end{align}
From Eqs.~\eqref{grmu} and \eqref{grnu}, we can extract the characteristic scales to minimize the large logarithms for the resummation of $\tilde{S}_{\mQ/\mQ}$,
\be
\mu_r \sim \frac{1}{\bar{b}},~~\nu_r \sim \frac{p_+}{m\mu_r} \sim \frac{p_+\bar{b}}{m}.  
\ee
Thus we see that $\nu_r$ has the same scaling as the large component of the residual momentum shown in Eq.~\eqref{resim}. 

As a consistency check between the $\mr{SCET_M}$ and bHQET calculations, we take the limit of $\tilde{D}_{\mQ/\mQ}(z,\bl{b})$ in Eq.~\eqref{DQQ1b} as $z$ goes to 1 with  power counting $mb \sim (1-z)^{-1}$. This result coincides with the combination of $C_{\mQ}$ and $S_{\mQ/\mQ}$ 
at NLO found above: 
\begin{align}
&\tilde{D}_{\mQ/\mQ}^{(1)} (z\to 1, \bl{b} \sim m^{-1}(1-z)^{-1},\mu,\nu ) 
= C_{\mQ}^{(1)} (m,\mu)  + \tilde{S}_{\mQ/\mQ}^{(1)} (z,\bl{b},\mu,\nu) \nnb \\
\label{consis} 
&~~ = \frac{\as C_F}{2\pi} \Biggl\{ \delta (1-z) \Bigl[2\ln\frac{\nu}{p_+}\cdot \ln \bar{b}^2\mu^2 + \frac{3}{2} \ln\frac{\mu^2}{m^2} +2 \Bigr] 
- \Bigl[\frac{2}{1-z} (1+2\ln(1-z))\Bigr]_+  \\
&~~+\frac{4}{(1-z)_+} \Bigl[K_0 ((1-z)mb) +\ln(1-z) \Bigr]
-2(1-z) \Bigl[\frac{bm}{1-z} K_1 ((1-z)mb) -\frac{1}{(1-z)^2} \Bigr] \Biggr\}\ . \nnb
\end{align}

\section{Full description on the heavy quark TMD fragmentation function}

\subsection{The TMD fragmentation function when $\qp \gg m$}
\label{TMDFFqmlm}

When $q_{\perp} \gg m$, the HQ TMDFF in Eq.~\eqref{defTMDFF} can be factorized due to this hierarchy of scales. To accomplish this, we need to first integrate out the fluctuations of $\qp^2$, then consider the fragmentation to the hadron at the lower scale $\mu \sim m$. 
Thus, the HQ TMDFF in this case can be matched onto the standard heavy quark fragmentation function~(HQ FF), which only depends on the longitudinal momentum fraction of the hadron. 
In  $\bb$-space, the factorization reads
\be 
\label{DHQfact}
\tilde{D}_{H/\mQ} (z,\bb;\mu,\nu) = \sum_{k} \int^1_z \frac{dx}{x} K_{k/\mQ} (x,\bb;\mu,\nu) D_{H/k}\bigl(\frac{z}{x},\mu \bigr)+\mO(mb),
\ee
where $D_{H/k} (z/x)$ is the standard FF to the heavy hadron $H$, and $k$ is the flavor of the fragmenting parton. Except for the case $k = \mQ$, the contributions from other partons are suppressed by at least $\as^2$ and can be ignored to the order we are considering. Note that since we are considering the limit $q_{\perp} \gg m$, here $mb \ll 1$.

From the NLO result of $\tilde{D}_{\mQ/\mQ}(z,\bb)$ in Eq.~\eqref{DQQ1b}, we can directly obtain the NLO result of $K_{\mQ/\mQ}$ in Eq.~\eqref{DHQfact} by matching onto the FF, $D_{\mQ/\mQ}(z/x)$. The result of Eq.~\eqref{DQQ1b} was been obtained with treatment of $mb \sim \mO(1)$, hence it can be considered as the full result in an expansion of $mb$. 
We must, therefore, extract the leading result from Eq.~\eqref{DQQ1b} in the limit $mb \to 0$. 
Accordingly, the following combinations of the Bessel functions in Eq.~\eqref{DQQ1b} can be expanded as 
\begin{align} 
\label{brelK0} 
K_0 \Bigl(\frac{1-z}{z}mb\Bigr) +\ln(1-z) &=  - \ln m \bar{b} + \ln z  + \mc{O}(mb), \\
\label{brelK1} 
\frac{bm}{1-z} K_1 \Bigl(\frac{1-z}{z}mb\Bigr)-\frac{z}{(1-z)^2} &= \frac{m^2b^2}{4z} \Bigl(-1+2\ln (1-z) m\bar{b}
\Bigr)+ \mc{O}(m^3b^3). 
\end{align}
The combination with $K_1$ can be safely ignored in this limit, and using Eq.~\eqref{brelK0} we obtain 
\begin{align} 
\label{DQQbs}
\tilde{D}_{\mQ/\mQ} \bigl(z,\bb\ll \frac{1}{m};\mu,\nu \bigr) &=1+ \frac{\as C_F}{2\pi} 
\Biggl\{ \delta(1-z)\cdot \Bigl(2\ln\frac{\nu}{p_+}+\frac{3}{2} \Bigr) \ln\bar{b}^2 \mu^2 \\
&\hspace{-2cm}-\Bigl[\frac{1+z^2}{1-z} \bigl(\ln m^2\bar{b}^2 (1-z)^2+1 \bigr)\Bigr]_+ + \Bigl(\frac{4z}{1-z}\Bigr)_+ \ln z + (1-z)(1+2\ln z)\Biggr\}\ . \nnb 
\end{align}

The NLO result for the standard HQ FF is well known,~\cite{Mele:1990cw} 
\be
\label{sDQQnlo}
D_{\mQ/\mQ} (z,\mu)    = \delta (1-z) + \frac{\as C_F}{2\pi} \Bigl[\frac{1+z^2}{1-z} \Bigl(\ln \frac{\mu^2}{m^2(1-z)} - 1\Bigr) \Bigr]. 
\ee
By subtracting this from the one-loop result of Eq.~\eqref{DQQbs} we obtain the one-loop result of $K_{\mQ/\mQ}$,\footnote{
This result is consistent with the result for the TMD beam function~\cite{Mantry:2009qz,Procura:2014cba}, which describes an incoming parton before a hard collision. The one-loop result for the TMD kernel of the beam function can be immediately obtained from the result of Eq.~\eqref{KQQ1} by replacing $\bar{b}/z \to \bar{b}$. 
Here the difference of $z$ is due to the fact the TMDFF describes the transverse momentum distribution of initiating parton before collinear splitting, while the beam function measures transverse momentum of  hard-colliding parton after the splitting. 
} 
\begin{align} 
\label{KQQ1}
K_{\mQ/\mQ}^{(1)} (z,\bb;\mu,\nu) &= \tilde{D}^{(1)}_{\mQ/\mQ} \bigl(z,\bb\ll \frac{1}{m};\mu,\nu \bigr) - D_{\mQ/\mQ}^{(1)} (z,\mu)  \\
&=\frac{\as C_F}{2\pi} \Bigl[\delta(1-z) \Bigl(2 \ln\frac{\nu}{p_+} + \frac{3}{2} \Bigr) \ln\bar{b}^2\mu^2  - \Bigl(\frac{1+z^2}{1-z} \Bigr)_+ \ln \frac{\bar{b}^2}{z^2} \mu^2 + 1-z \Bigr]\ . 
\nnb 
\end{align}
As expected, the TMD kernel $K_{\mQ/\mQ}$ does not depend on the heavy quark mass $m$ and its characteristic scale is of order $\mu \sim 1/\bar{b}$. 

\subsection{Nonperturbative contribution to the fragmentation function when $\qp \gg \Lambda_{\rm QCD}$} 
\label{TMDFFqgglambda}

Thus far we have not considered the hadronization effects governed by nonperturbative physics at scale $\Lambda_{\mr{QCD}}$. For a heavy-light hadron $H$ involving a heavy quark, like a $B$ meson, the hadronization in the fragmenting process is through low energy interactions of the heavy quark,  adequately described by bHQET. Further, in bHQET the interactions are entirely mediated by the residual gluon that carries only a small fraction of the energy. Hence the fragmenting process for hadronization dominantly occurs in the large-$z$ region where the heavy quark in the final state carries most of the energy in the process.  

To include the nonperturbative contribution, the standard HQ FF can be written as~\cite{Nason:1999zj,Cacciari:2005uk,Fickinger:2016rfd} 
\be 
\label{stFF} 
D_{H/i}(x,\mu) = \int^1_x \frac{dz}{z} D_{\mQ/i} \Bigl(\frac{x}{z},\mu\Bigr) \phi_{H/\mQ} (z),  
\ee
where $D_{Q/i}$ is the FF at the parton level to be computed perturbatively and $\phi_{H/\mQ}$ is the nonpertubative piece describing the modification due to hadronization. 
The distribution $\phi_{H/\mQ}$ is strongly peaked in the large-$z$ region. When the $x$ in $D_{H/i}(x)$ probes the region far away from the endpoint, i.e., $1-x \sim \mO(1)$, the nonperturbative contribution should be negligible since $m(1-x) \gg \Lambda_{\rm QCD}$, hence we guess that $\phi_{H/\mQ}$ acts like a delta function~\cite{Fickinger:2016rfd}, 
\be 
\phi_{H/\mQ} (z) \approx N_H \delta (1-z), 
\ee
where $N_H$ is the nonperturbative fractional parameter for the hadronization. In bHQET, $N_H$ is defined by~\cite{Fickinger:2016rfd} 
\be 
N_H = \frac{1}{4N_c m_H } \sum_{X_r} \mr{Tr} \langle 0 | Y_{\n}^{r\dagger} h_n | H_v X_r \rangle 
\langle H_v X_r | \bar{h}_n Y_{\n}^r \nn | 0 \rangle, 
\ee
where $|H_v \rangle = |H \rangle/\sqrt{m_H}$. When we consider the sum over all the hadrons containing the heavy quark, 
it should satisfy $\sum_H N_H =1$.

In Eq.~\eqref{stFF}, the NLO perturbative result for $D_{\mQ/\mQ}$ was shown in Eq.~\eqref{sDQQnlo}, while the result for $D_{\mQ/g}$ reads~\cite{Mele:1990cw} 
\begin{align}
D_{\mQ/g}  (z,\mu) & = \frac{\as}{2\pi} \frac{z^2 + (1-z)^2}{2} \ln \frac{\mu^2}{m^2}\ . 
\end{align}
As $z$ approaches 1, $D_{\mQ/\mQ}$ dominates over  $D_{\mQ/g}$ and, similarly to Eq.~\eqref{HQFFF}, it factorizes 
\be
\label{stfff} 
D_{\mQ/\mc{Q}} (z\to 1,\mu) = C_{\mQ} (m,\mu) S_{\mQ/\mQ} (z,\mu),  
\ee
where $C_{\mQ}$ was introduced in Eq.~\eqref{CQnlo}, and $S_{\mQ/\mQ}(z)$ at NLO is given by 
\cite{Neubert:2007je,Fickinger:2016rfd} 
\begin{align} 
\label{sznlo} 
S_{\mQ/\mQ} (z,\mu) = \delta(1-z) +  \frac{\as C_F}{2\pi} \Biggl\{&\delta(1-z) \Bigl(\ln\frac{\mu^2}{m^2} - \frac{1}{2}\ln^2\frac{\mu^2}{m^2} -\frac{\pi^2}{12} \Bigr) \\
& + \Bigl[\frac{2}{1-z} \Bigl(\ln\frac{\mu^2}{m^2(1-z)^2} - 1 \Bigr) \Bigr]_+ \Biggr\}\ . \nnb
\end{align} 

When we consider nonperturbative implications for the HQ TMDFF with $\qp \gg \Lambda_{\mr{QCD}}$, we can basically apply the same approach as Eq.~\eqref{stFF}, hence we will employ the same nonperturbative function. As a result, when $\qp \gg \Lambda_{\mr{QCD}}$, the HQ TMDFF can be written as 
\be 
\label{npimp} 
D_{H/i} (x, \qp, \mu,\nu)  = \int^1_x \frac{dz}{z} D_{\mQ/i} (z,\qp,\mu,\nu) \phi_{H/\mQ} (z). 
\ee
Here the NLO result of $D_{\mQ/i=\mQ}$ was obtained in Eq.~\eqref{DQQ1}, and the one loop result of $D_{\mQ/g}$ is 
\be 
\label{DQgtmd}
D_{\mQ/g} (z,\qp,\mu) = \frac{\as T_F}{2\pi} \frac{1}{\qps+m^2/z^2} \Bigl[1-2z(1-z)\frac{\qps}{\qps+m^2/z^2}\Bigr],
\ee
where $T_F = 1/2$.

When $\qp$ is much smaller than the heavy quark mass $m$, $D_{\mQ/\mQ}$  dominates over $D_{\mQ/g}$ and, as shown in Eq.~\eqref{HQFFF}, $D_{\mQ/\mQ}$ can be additionally factorized 
as\footnote{Through comparison of Eq.~\eqref{HQFFF} with Eq.~\eqref{npimp} and Eq.~\eqref{HQFFFpl}, we  can relate 
\be
S_{H/\mQ}(x,\qp\gg \Lambda_{\rm QCD},\mu,\nu) = \int^1_x \frac{dz}{z} S_{\mQ/\mQ} (z,\qp,\mu,\nu) \phi_{H/\mQ} (x/z). 
\ee
} 
\be 
\label{HQFFFpl} 
D_{\mQ/\mc{Q}} (z,\blp{q}\ll m,\mu,\nu) = C_{\mQ} (m,\mu) S_{\mQ/\mQ} (z,\blp{q},\mu,\nu). 
\ee
We have also discussed the HQ TMDFF for $\qp \gg m$ in subsection~\ref{TMDFFqmlm}. As shown there, the TMDFF can be matched onto the standard FF with the fluctuations of $\qp^2$ integrated out. Therefore, the nonperturbative piece can be genuinely included in the standard FF as in Eq.~\eqref{stFF}.

For the parameterization of the nonperturbative FF, $\phi_{H/\mQ}$, 
%shown in eq.~\eqref{npimp} (also in eq.~\eqref{stFF}), 
we adopt the model introduced in Ref.~\cite{Neubert:2007je,Fickinger:2016rfd}, 
\be 
\label{npFF} 
\phi_{H/\mQ} (z) = N_H \frac{m}{\lambda_H} \frac{(p+1)^{p+1}}{\Gamma(p+1)} \Bigl(\frac{m}{\lambda_H} (1-z) \Bigr)^p 
e^{-(p+1) (1-z)m/\lambda_H}.
\ee
This was originally introduced in momentum space in $\hat{\omega} = (1-z)mv_+$, where $\phi_{H/Q} (\hat{\omega}) = \phi_{H/\mQ} (z) \cdot |dz/d\hat{\omega}|$. The integral over the full range of $\hat\omega$ is normalized to unity,\footnote{Throughout this paper, we do not specify the heavy hadron but include all the possible heavy-light hadrons. Hence $N_H$ is given by one. 
}
\be
\int^{\infty}_0 d\hat{\omega} ~\phi_{H/Q} (\hat{\omega})  =1.  
\ee
$\lambda_H$ in Eq.~\eqref{npFF} is a quantity of order $\Lambda_{\rm QCD}$ and is related to the first moment of $\phi_{H/Q} (\hat{\omega})$,
\be 
\int^{\infty}_0 d\hat{\omega}~ \hat{\omega} \phi_{H/Q} (\hat{\omega}) = \frac{\lambda_H}{v_+}\ .
\ee
One advantage of using Eq.~\eqref{npFF} is that, in the limit $m/\lambda_H \to \infty$, the nonperturbative FF becomes $\phi_{H/\mQ} \approx 
N_H \delta(1-z)$. So, as long as $m/\lambda_H$ is a large value much greater than 1, the nonperturbative effects predominantly make an impact on the endpoint region with $1-z \sim \mO(\Lambda/m)$. Away from the endpoint, the nonperturbative effects are small.  

\subsection{Summary: the HQ TMDFF with $\qp \gg \Lambda_{\rm{QCD}}$}

In this subsection, we summarize our results of the HQ TMDFF with the different hierarchies between $\qp$ and $m$. With the assumption that $\qp \gg \Lambda_{\rm{QCD}}$, we can describe the transverse-momentum dependent part purturbatively and can put the nonperturbative effects fully into $\phi_{H/\mQ}$. 
Here we show the TMDFFs in the $\bb$-space comparing the sizes of $b$ and $1/m$: 

\begin{enumerate}[label=\roman*)] 
\item 
$b \ll 1/m~(q_{\perp} \gg m)$
\be
\tilde{D}_{H/i} (z,\bb,\mu,\nu) = \sum_{j} \int^{1}_z \frac{dx}{x} K_{j/i}(x,\bb,\mu,\nu) D_{H/j} \bigl(\frac{z}{x},\mu\bigr)+\mO(mb), 
\ee
where, from Eq.~\eqref{stFF}, 
\be
D_{H/j} \bigl(\frac{z}{x},\mu) = \int^1_{z/x} \frac{dy}{y} D_{\mQ/j} (y,\mu) \phi_{H/\mQ} \bigl(\frac{z}{xy}\bigr).
\ee
Here the one loop result of the TMD kernel $K_{\mQ/\mQ}$ was presented in Eq.~\eqref{KQQ1}. We also computed the one-loop results for the kernels with other flavors, which are
\begin{align} 
\label{Kgq1}
K^{(1)}_{g/q} (z,\bb,\mu) &= \frac{\as C_F}{2\pi} \Bigl[-\frac{1+(1-z)^2}{z} \ln\frac{\bar{b}^2 \mu^2}{z^2} +z\Bigr]\ , \\ 
\label{Kqg1}
K^{(1)}_{q/g} (z,\bb,\mu) &= \frac{\as T_F}{2\pi} \Bigl[-\bigl(z^2+(1-z)^2) \ln \frac{\bar{b}^2 \mu^2}{z^2} - 2z(1-z) \Bigr]\ , \\
\label{Kgg1}
K^{(1)}_{g/g} (z,\bb,\mu,\nu) &= \frac{\as C_A}{2\pi} \Bigl\{ \delta(1-z) \cdot 2\ln\frac{\nu}{p_+} \ln \bar{b}^2\mu^2 \\
&~~~-2\Bigl[\frac{z}{(1-z)_+} + \frac{1-z}{z} + z(1-z) \Bigr] \ln\frac{\bar{b}^2 \mu^2}{z^2}
\Bigr\}\ . \nnb 
\end{align}

\item 
$b \sim 1/m~(q_{\perp} \sim m)$
\be 
\label{tDbs}
\tilde{D}_{H/i} (z,\bb,\mu,\nu) = \int^{1}_z \frac{dx}{x} \tilde{D}_{\mQ/i} (x,\bb,\mu,\nu) \phi_{H/\mQ} \bigl(\frac{z}{x}\bigr).
\ee
Here the NLO result of $\tilde{D}_{\mQ/\mQ} (z,\bb)$ was obtained in Eq.~\eqref{DQQ1b}, and the Fourier transform of Eq.~\eqref{DQgtmd}, $\tilde{D}_{\mQ/g}$, is given by 
\be
\tilde{D}_{\mQ/g} (z,\bb,\mu) = \frac{\as T_F}{\pi} \Bigl[(z^2+(1-z)^2) K_0 \bigl(\frac{mb}{z}\bigr) - (1-z)mb K_1 \bigl(\frac{mb}{z}\bigr)\Bigr]\ . 
\ee

\item 
$1/\Lambda_{\rm QCD} \gg b\gg 1/m~(m\gg q_{\perp} \gg \Lambda_{\rm QCD})$ 

In this case, $\tilde{D}_{H/i} (z,\bb)$ is approximated by $\tilde{D}_{H/\mQ}(z,\bb)$, and $\tilde{D}_{\mQ/\mQ}(z,\bb)$ can be refactorized to $C_{\mQ} (m) \cdot \tilde{S}_{\mQ/\mQ} (z,\bb)$, which is the Fourier transform of Eq.~\eqref{HQFFFpl}. Therefore, 
\be 
\label{tDbl}
\tilde{D}_{H/i} (z,\bb,\mu,\nu) = C_{\mQ}(m,\mu) \int^{1}_z \frac{dx}{x} \tilde{S}_{\mQ/\mQ} (x,\bb,\mu,\nu) \phi_{H/\mQ} \bigl(\frac{z}{x}\bigr)+\mO\bigl(\frac{1}{mb}\bigr), 
\ee 
where the NLO results of $C_{\mQ}$ and $\tilde{S}_{\mQ/\mQ}$ were presented in Eqs.~\eqref{CQnlo} and \eqref{SQb1}, respectively. 
\end{enumerate}

\section{The heavy quark TMD jet fragmentation function} 
\label{HQTMDJFF}

In this section, as an application of the HQ TMDFF, we will analyze heavy quark fragmentation inside a given observed jet 
constructing the factorization theorem for the HQ TMD jet fragmentation function (JFF). 
By considering the TMD fragmenting process within a jet, we can closely delineate the substructure of a jet involving the heavy quark and acquire direct or useful information on the hadronization of the heavy quark. 

\begin{figure}[h]
\begin{center}
\includegraphics[height=4cm]{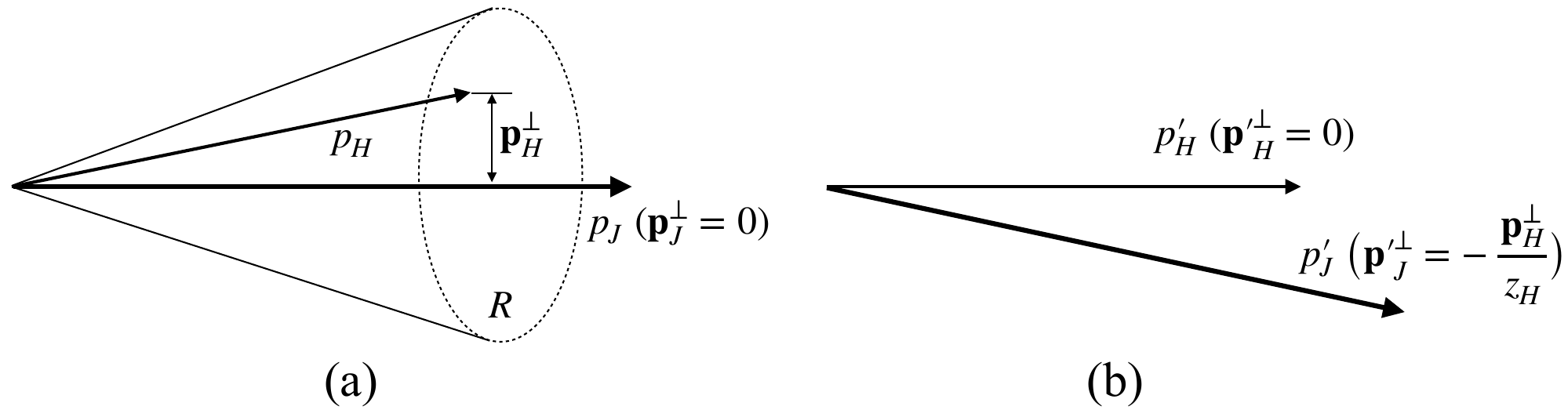}
\end{center}
\vspace{-0.5cm}
\caption{\label{fig3}
(a): Description of fragmentation to the hadron inside a jet with radius $R$ in the jet frame, where the transverse momentum of the jet is set to zero. 
(b): Description of the fragmentation in the hadron frame. Here the momentum fraction, $z_H=p_H^+/p_J^+$, is the same in both frames.  
}
\end{figure}

As illustrated in Fig.~\ref{fig3}(a), we consider the transverse momentum distribution of the hadron with respect to the standard jet axis, which lies along the total momentum of the jet. For a jet with small radius $R$, the typical jet size is $E_J R$ for $e^+e^-$ annihilation or $p_J^T R$ for hadronic collisions, where $p_J^T$ is the large transverse momentum relative to the beam axis. 
Since we are considering the small transverse momentum distribution of the hadron relative to the jet axis, $\blpu{p}_H$, we will assign the limit $\blpu{p}_H \ll E_JR,~p_J^TR$. 
Then the transverse motion of the hadron is described by  collinear and  collinear-soft (csoft) interactions, whose momenta scale as 
\bea
\label{colsca} 
p_c^{\mu} &=& (p_c^+,p_c^-,\blpu{p}_c)  \sim  (E_J, p_H^{\perp 2}/E_J, p_H^{\perp}), \\
\label{csoftsca}
p_{cs}^{\mu}  &=& (p_{cs}^+,p_{cs}^-,\blpu{p}_{cs}) \sim  (p_H^{\perp}/R, p_H^{\perp} R, p_H^{\perp}), 
\eea
where $p_H^{\perp} \equiv |\blpu{p}_H |$. 
Note that csoft interactions can discern the jet boundary, while collinear interactions cannot.

In building up the factorization theorem, as illustrated in Fig~\ref{fig3}(b), it is convenient to consider this fragmentation process in the hadron frame because the factorization usually describes TMD behaviors of  the (collinear/csoft) overall initiating partons. In the hadron frame the total transverse momentum inside the jet is nonzero, and is related to the transverse momentum of the hadron in the jet frame that is an observable in the experiment by
\be
\label{relpp}
\qp\equiv\blpu{p'}_J = \sum_{i\in J} \blpu{p'}_{c,i}  + \sum_{j\in J} \blpu{p'}_{cs,j} = - \frac{\blpu{p}_H}{z_H}.  
\ee
Here $\blpu{p}_H$ is the momentum in the jet frame and we denoted the momenta in the hadron frame with primes.\footnote{Since we do not consider the limit $z_H \ll 1$, throughout this paper both the transverse momenta $\qp$ and $\blpu{p}_H$ are power counted as having the same scaling.} 

\subsection{The TMD JFF module} 

In this subsection, focusing on the inclusive jet production for $e^+e^-$ annihilation we consider the following differential jet cross section to observe a hadron inside the jet: 
%\be 
%\label{defJFF} 
%F_J (z_H, \blpu{p}_H ; E_J, R) = \frac{d\sigma (e^+e^- \to J(H) X)}{dE_J dz_H d^2 \blpu{p}_H} \Biggl / 
%\frac{d\sigma (e^+e^- \to JX) }{dE_J}\ , 
%\ee
\be 
\label{difsig}
\frac{d\sigma (e^+e^- \to J(H) X)}{dE_J dz_H d^2 \blpu{p}_H}\ , 
\ee
where $J(H)$ denotes the jet that includes a hadron $H$, $z_H = p_H^+/p_J^+$ is the hadron momentum fraction over the jet, and  $\blpu{p}_H$ is the hadron transverse momentum with respect to a jet axis. 
If we divide Eq.~\eqref{difsig} by $d\sigma/dE_J$, we obtain a probability finding a hadron with $z_H$ and $\blpu{p}_H$ inside a jet with $E_J$ (TMD JFF). 
In hadron collisions, we can similarly define the differential cross section with respect to $p_T^J$ (as well as rapidity) rather than $E_J$. 
For clustering a jet, we consider the anti-$\mr{k_T}$ algorithm~\cite{Cacciari:2008gp,Ellis:2010rwa}. 
If the jet radius $R$ is small, the cross sections in Eq.~\eqref{difsig} are factorized into the hard and jet parts. For the jet containing $H$ we have 
\bea  
\frac{d\sigma (e^+e^- \to J(H) X)}{dE_J dz_H d^2 \blpu{p}_H} 
&=& \frac{d\sigma (e^+e^- \to J(H) X)}{z_H^2 dE_J dz_H d^2 \qp} \nnb \\
\label{dsfac}
&=& \sum_k \int^1_{x_J} \frac{dx}{x} \frac{d\hat{\sigma}}{dE_k} \left(\frac{x_J}{x}, \mu\right)   
\cdot \frac{1}{z_H^2} \mc{G}_{J(H)/k} (x,z_H,\qp, E_J, R,\mu) 
\eea
where $\hat{\sigma}$ is the partonic cross section and $\mc{G}_{J(H)/k}$ is the semi-inclusive TMD fragmenting jet function from  parton $k$ to  hadron $H$ inside the jet $J$. This formalism has been applied to jet production with massless partons~\cite{Kang:2017glf}. 
As denoted in Eq.~\eqref{relpp}, $\qp$ is the jet transverse momentum in the hadron frame and  
the longitudinal momentum variables in Eq.~\eqref{dsfac} are 
\be
x_J = \frac{2 p_{\rm tot}\cdot p_J}{p_{\rm tot}^2} = \frac{2E_J}{Q} \sim \frac{p_J^+}{Q},~~x=\frac{p_J^+}{p_k^+},~~z_H=\frac{p_H^+}{p_J^+}\ ,    
\ee
where $p_{\rm tot}$ is the total momentum of the incoming electron and positron, and $p_{\rm tot}^2 = Q^2$.
Here the parton $k$, the jet $J$, and the hadron $H$ are all described to be collinear in the $n$-direction.

Since we are taking the limit $E_J R \gg \qp$, the jet function $\mc{G}_{J(H)/k}$ can be further factorized. 
In this case it is useful to express the factorization using the fragmentation function to a jet (FFJ)~\cite{Kang:2016mcy,Dai:2016hzf}. 
To NLO in $\as$, we can refactorize $\mc{G}_{J(H)/k}$ as~\cite{Dai:2016hzf,Dai:2018ywt} 
\be
\label{refac}
\mc{G}_{J(H)/k} (x,z_H,\qp, E_J, R,\mu) =\sum_l D_{J_l/k} (x,E_J R,\mu) \Phi_{H/J_l} (z_H,\qp; E_J, R).
\ee
Here $D_{J_l/k}$ is the FFJ from  parton $k$ to $J_l$, where $J_l$ indicates the jet initiated by parton $l$. 
Beginning at NNLO in $\as$, Eq.~\eqref{refac} does not hold due to the presence of $1\to 3$ splitting processes. 
However, 
this refactorization is advantageous to understanding the jet substructure and the fragmentation process within the jet.
Note that the combination of $D_{J_l/k}$ and $d\hat{\sigma}/dE_k$ together is scale-invariant. Hence the remaining function $\Phi_{H/J_l}$ must be also scale-invariant. Moreover, $\Phi_{H/J_l}$ can be normalized to 
\be
\label{JFFnorm} 
\sum_H \int dz_H z_H  \int_{J} d^2\qp~\Phi_{H/J_l} (z_H, \qp) =1,  
\ee
where the integration region for $\qp$ is limited to be inside the jet. From now we will call $\Phi_{H/J_l}$ ``the JFF module", which is responsible for the hadron fragmentation and its jet substructure. 

\subsection{Factorization of the heavy quark TMD JFF module} 
\label{factmod} 

Since we are  interested in HQ TMD fragmentation, in this section we consider the factorization of the HQ TMD JFF module, $\Phi_{H/J_\mQ} (z_H, \qp;E_J,R,m)$, where we take the heavy quark mass to be $m \ll E_JR$. 
With the hierarchy $E_J R \gg \qp, m$, the JFF module $\Phi_{H/J_\mQ}$ can fully include the HQ TMDFF $D_{H/\mQ}$. 
Furthermore, as introduced in Eq.~\eqref{csoftsca}, the csoft interaction enters to describe the transverse motion of the hadron within a jet. 
Finally, from Eq.~\eqref{JFFnorm}, the JFF module has the normalization factor, which is obtained by integrating over the full phase space inside a jet. 

As a result, we present the factorization theorem for $\Phi_{H/J_\mQ}$, 
\bea 
\label{JFFfac} 
\Phi_{H/J_\mQ} (z_H, \qp ;E_J,R,m) &&  \\
&&\hspace{-4cm}=H_J (E_J R,\mu) \int d\blp{k}^2 d\blp{l}^2 S_R (\blp{l},\mu,\nu) D_{H/\mQ} (z_H,\blp{k},m,\mu,\nu) \delta^{(2)}(\blp{k}+\blp{l}-\qp). \nnb 
\eea
Here $H_J$ is the hard-collinear function governed by the typical jet scale $E_J R$, $S_R$ is the TMD csoft function,
% for the interaction scaling as Eq.~\eqref{csoftsca}, 
and $D_{H/\mQ}$ is the HQ TMDFF introduced in Eq.~\eqref{defTMDFF}. 

Since $H_J$ is the normalization factor for integrating over the full phase space within a jet, it is given by the inverse of the heavy quark integrated jet function~\cite{Dai:2018ywt}, 
\be 
H_J(E_J R, m, \mu) = \mc{J}^{-1}_{\mQ} (E_J R,m,\mu).
\ee
In the limit we are considering, $E_J R \gg m$, the heavy quark mass $m$ can be safely ignored. We can therefore use the result of the integrated jet function for massless quarks~\cite{Cheung:2009sg,Ellis:2010rwa,Liu:2012sz,Chay:2015ila}, and so $H_J$ at NLO in $\as$ is
\bea
H_J(E_J R \gg m, \mu) &\approx& \mc{J}^{-1}_q (E_JR,\mu)  \nnb \\
\label{HJnlo} 
&=& 1- \frac{\as C_F}{2\pi} \Bigl(\frac{3}{2} \ln\frac{\mu^2}{E_J^2 R^2 }
+\frac{1}{2} \ln^2 \frac{\mu^2}{E_J^2 R^2} +\frac{13}{2} -\frac{3\pi^2}{4}\Bigr).  
\eea

The csoft function $S_R$ consists of the decoupled csoft Wilson lines from collinear sectors, given by  
\be 
\label{defSR} 
S_R (\lp,\mu,\nu) = \frac{1}{N_c} \mr{Tr}~\langle 0 | \tilde{Y}_{n,cs} Y_{\n,cs}^{\dagger} \delta^{(2)} (\lp + \Theta_{in} \cdot  \bsp{\mP})  Y_{\n,cs} \tilde{Y}_{n,cs}^{\dagger} |0\rangle, 
\ee 
where $N_c$ is the number of colors, and $\Theta_{in}\cdot \bsp{\mP}$ is the derivative operator taking transverse momentum only when a gluon radiates inside a jet.   
$\tilde{Y}_{n,cs}$ and $Y_{\n,cs}$ are csoft Wilson lines. The  tilded Wilson line~\cite{Chay:2004zn} has a different path compared with the standard Wilson line, 
\be 
\tilde{Y}_{n,cs}^{\dagger} (x) = \mr{P} \exp \Bigl[ig \int^{\infty}_x ds n\cdot A_{cs} (n s)\Bigr],  
\ee
where `P' represents path ordering. 

The one-loop result for the TMD csoft function was obtained in Ref.~\cite{Bain:2016rrv,Kang:2017mda,Kang:2017glf}.  
We also illustrate the calculation in Appendix~\ref{TMDcsoft}. 
To NLO in $\as$, the renormalized csoft function is 
\bea
\label{SRnlom} 
S_R(\lp,\mu,\nu) &=& \frac{1}{\pi}\delta(\lps) + \frac{\as C_F}{2\pi^2} \Biggl\{\delta(\lps) \Bigl(
 -2 \ln \frac{\mu^2}{\Lambda^2}\ln\frac{\nu R}{2\Lambda}
+\frac{1}{2} \ln^2 \frac{\mu^2}{\Lambda^2} - \frac{\pi^2}{12} \Bigr)  \\
&&\phantom{\frac{1}{\pi}\delta(\lps) + \frac{\as C_F}{2\pi^2}\Biggl\{ }
+\Bigl[\frac{1}{\lps} \ln \frac{\nu^2 R^2}{4 \lps} \Bigr]_{\Lambda^2} \Biggr\}. \nnb 
\eea
In $\bb$-space, it is given by 
\bea 
\tilde{S}_R (\bl{b},\mu,\nu) &=& \int d^2 \lp e^{i\bl{b}\cdot \lp}  S_R(\lp,\mu,\nu) \nnb \\
\label{SRnlob} 
&=& 1+ \frac{\as C_F}{2\pi} \Biggl(-\ln\bar{b}^2\mu^2 \ln \frac{\nu^2 R^2}{4\mu^2} - \frac{1}{2} \ln^2\bar{b}^2\mu^2 - \frac{\pi^2}{12} \Biggr).
\eea
From this result we understand that the characteristic csoft scales are
\be 
\mu_{cs} \sim 1/\bar{b} \sim \kp,~~\nu_{cs} \sim \frac{\mu}{R/2} \sim \frac{\kp}{R/2} \sim p_{cs}^+\ . 
\ee 
Note that the characteristic rapidity scale for the csoft function corresponds to the largest momentum component of the csoft momentum. 
Hence, as we will see later, when combined with $D_{H/Q}$ in Eq.~\eqref{JFFfac},  the evolution of the rapidity scale between $\nu_c$ and $\nu_{cs}$ will resum the  large logarithm 
\be 
\ln \frac{\nu_c}{\nu_{cs}} \sim \ln \frac{p_c^+}{p_{cs}^+} \sim \ln \frac{E_J R}{q_{\perp}}\ .  
\ee 

For convenience for the eventual running, we express the factorization theorem for $\Phi_{H/J_\mQ}$ in Eq.~\eqref{JFFfac} in  $\bl{b}$-space, 
\bea 
\tilde{\Phi}_{H/J_\mQ} (z_H, \bl{b};E_J,R,m) &=& \int d^2 \qp e^{i\bl{b}\cdot \qp} \Phi_{H/J_\mQ} (z_H, \qp;E_J,R,m) \nnb \\
\label{JFFfacb} 
&=& H_J (E_J R,\mu) \tilde{S}_R (\bl{b},\mu,\nu) \tilde{D}_{H/\mQ} (z_H,\bl{b},m,\mu,\nu), 
\eea 
where, for $\tilde{D}_{H/\mQ}$, the NLO result at the parton level (i.e., $\tilde{D}_{\mQ/\mQ}$) is shown in Eq.~\eqref{DQQ1b}.   
Combining the one-loop results for all the factorized functions in Eq.~\eqref{JFFfacb}, 
we can easily check that $\tilde{\Phi}_{H/J_\mQ}$ is independent of the factorization scales, $\mu$ and $\nu$, with the parton-level result 
\bea 
\tilde{\Phi}_{\mQ/J_\mQ} (z, \bl{b};E_J,R,m) = 1+\frac{\as C_F}{2\pi} \Biggl(-2 \ln^2 (\bar{b} E_J R) + 3 \ln (\bar{b} E_J R) + \ln(\bar{b} m) + \cdots \Biggr),  
\eea 
where we have suppressed the non-logarithmic terms at NLO. 

As investigated in Sec.~\ref{smallq}, when $q_{\perp} \ll m$, the HQ TMDFF can be additionally factorized as shown in Eq.~\eqref{HQFFF}. We have, for  $\tilde{\Phi}_{H/J_\mQ} (z_H,\bl{b} \gg 1/m)$,
\bea 
&&\tilde{\Phi}_{H/J_\mQ} (z_H \to 1, \bl{b} \sim (m(1-z_H))^{-1} ;E_J,R,m) \nnb \\
\label{JFFfacbh}
&&~~~~~~~~~~~
= H_J (E_J R,\mu) C_{\mQ} (m, \mu) \tilde{S}_R (\bl{b},\mu,\nu) \tilde{S}_{H/\mQ} (z_H,\bl{b},m,\mu,\nu). 
\eea 
In this case the contributions are dominated by the large $z_H$ region. 
If $\bb \ll 1/\Lambda_{\rm QCD}$, $\tilde{S}_{H/\mQ}$ can be given by the convolution of $\tilde{S}_{\mQ/\mQ}$ and $\phi_{H/\mQ}$ as shown in Eq.~\eqref{tDbl}. 
For NLO result of $\tilde{S}_{\mQ/\mQ}$ at parton level is shown in Eq.~\eqref{SQb1}.

\subsection{Resummation of the heavy quark TMD JFF module: purturbative results} 

In this subsection, we investigate resummation of the large logarithms (except nonglobal logarithms) in the HQ TMD JFF module in the perturbative limit to  next-to-leading logarithmic (NLL) accuracy.
%\footnote{
%For the complete resummation to NLL accuracy, we have to include the contribution from large nonglobal logarithms, which in our case arise from the factorization between $H_J$ and $\tilde{S}_R$ in which the relevant modes can recognize the jet boundary.  In the limit $E_J R \gg m$ we consider, the heavy quark mass effects can be safely ignored in $H_J$ and $\tilde{S}_R$, hence the contribution becomes the same as the case of a light quark. The same discussion holds for TMD distribution with respect to thrust axis which is analyzed in section~\ref{thrust}. 
%For the resummation of the nonglobal logarithms for the massless case we refer to Ref.~\cite{Kang:2020yqw}. 
%}
For this we consider the TMD JFF module at parton level, i.e, $\Phi_{\mQ/J_{\mQ}}$. In resumming, it is convenient to use 
the factorized result in  $\bl{b}$-space shown in Eq.~\eqref{JFFfacb}. Then, after Fourier transforming, the TMD module in  momentum space is 
\bea 
&&\Phi_{\mQ/J_{\mQ}} (z_H, \qp ;E_J,R,m) = \int \frac{d^2 \bl{b}}{(2\pi)^2} e^{-i\qp\cdot \bl{b}} 
\tilde{\Phi}_{\mQ/J_{\mQ}} (z_H, \bl{b} ;E_J,R,m) \nnb \\
\label{TMDJFFfact} 
&&\hspace{1cm}= H_J (E_J R,\mu_f) \int \frac{db}{2\pi} b J_0 (b|\qp|) 
\cdot \tilde{S}_R (\bl{b},\mu_f,\nu_f) \tilde{D}_{\mQ/\mQ} (z_H,\bl{b},m,\mu_f,\nu_f), 
\eea 
where $J_0$ is the Bessel function of the first kind. $\mu_f$ and $\nu_f$ are the factorization scales. 
These factorization scales can be set arbitrarily since their overall dependences cancel in the TMD module. 
In Eq.~\eqref{TMDJFFfact}, large logarithms in each factorized function can be automatically resummmed through renomalization 
group (RG) evolutions from the characteristic scales to the factorization scales, $(\mu_f,\nu_f)$. 

For the complete resummation to NLL accuracy, we have to include  contributions from large nonglobal logarithms, which in our case arise from the factorization between $H_J$ and $\tilde{S}_R$ in which the relevant modes can recognize the jet boundary.  In the limit $E_J R \gg m$ we consider, the heavy quark mass effects can be safely ignored in $H_J$ and $\tilde{S}_R$, hence the contribution is the same as the case of a light quark. The same discussion holds for TMD distribution with respect to the thrust axis that is analyzed in section~\ref{thrust}. In this paper, we only consider resummation of large global logarithms based on the factorization theorem in Eq.~\eqref{TMDJFFfact}. For the resummation of nonglobal logarithms, we refer to Ref.~\cite{Kang:2020yqw}, of which the result for a light quark can be also applied to our case as long as $E_JR \gg m$.

The anomalous dimension for the evolution of $H_J$ is given by 
\be 
\label{gammaH} 
\gamma_H = \frac{1}{H_J} \frac{dH_J}{d\ln\mu} = - \Gamma_C (\as) \ln\frac{\mu^2}{E_J^2 R^2} + \hat{\gamma}_H (\as), 
\ee
where $\Gamma_C(\as)$ is the cusp anomalous dimension~\cite{Korchemsky:1987wg,Korchemskaya:1992je}. When expanded as $\sum_{k=0} \Gamma_{k}(\as/4\pi)^{k+1}$, the first two coefficients are given by 
\be
\Gamma_{0} = 4C_F,~~~\Gamma_{1} = 4C_F \Bigl[\bigl(\frac{67}{9}-\frac{\pi^2}{3}\bigr) C_A - \frac{10}{9} n_f\Bigr].
\ee
The non-cusp part of $\gamma_H$ in Eq.~\eqref{gammaH} is $\hat{\gamma}_H = -3 \as C_F /(2\pi)+\mc{O}(\as^2)$.    

The anomalous dimensions for $\mu$- and $\nu$-evolutions of $\tilde{S}_R$ are respectively given by 
\bea 
\label{gamSmu} 
\tilde{\gamma}_{\tilde{S}}^{\mu} &=& \frac{1}{\tilde{S}} \frac{d\tilde{S}}{d\ln\mu}  = \Gamma_{C} (\as) \ln \frac{4\mu^2}{\nu^2 R^2} + \hat{\gamma}_{\tilde{S}}, \\ 
\label{gamSnu} 
\tilde{\gamma}_{\tilde{S}}^{\nu} &=& \frac{1}{\tilde{S}} \frac{d\tilde{S}}{d\ln\nu}  = -2 a_{\Gamma} (\mu,1/\bar{b}),
\eea
where $\hat{\gamma}_{\tilde{S}} = \mO(\as^2)$, and the function $a_{\Gamma}$ is 
\be
a_{\Gamma} (\mu_1,\mu_2) = \int^{\mu_1}_{\mu_2} \frac{d\mu}{\mu} \Gamma_C (\as(\mu)).
\ee
Equations~\eqref{gamSmu} and \eqref{gamSnu} should satisfy the relation, 
\be 
\frac{d }{d\ln\nu} \gamma_{\tilde{S}}^{\mu}= \frac{d}{d\ln\mu} \gamma_{\tilde{S}}^{\nu}\ . 
\ee 
The anomalous dimensions for $\tilde{D}_{\mQ/\mQ}$ at leading order in $\as$ have been introduced in Eqs.~\eqref{gfmu} and \eqref{tgQnu}. 
To NLL accuracy, they read 
\begin{align}
\label{gamQmu} 
\tilde{\gamma}_{\mQ}^{\mu} &= %\frac{1}{\tilde{D}_{\mQ/\mQ}} \frac{d\tilde{D}_{\mQ/\mQ}}{d\ln\mu}  = 
\Gamma_{C} (\as) \ln \frac{\nu^2}{(2E_J)^2} + \hat{\gamma}_{\mQ}, \\ 
\label{gamQnu} 
\tilde{\gamma}_{\mQ}^{\nu} &= 
%\frac{1}{\tilde{D}_{\mQ/\mQ}} \frac{d\tilde{D}_{\mQ/\mQ}}{d\ln\nu}  = 
2 a_{\Gamma} (\mu,1/\bar{b}),    
\end{align}
where $\hat{\gamma}_{\mQ}=3\as C_F/(2\pi)+\mO(\as^2)$.  

Solving RG equations for the anomalous dimensions, we can systematically resum and exponentiate the large logarithms in Eq.~\eqref{TMDJFFfact}. If we consider the evolution over $\mu$ with the rapidity scale fixed, the exponentiation factor is  
\bea 
&&\ln U(\mu_f,\mu_h,\mu_c,\mu_{cs};\nu,\nu') = \ln U_H (\mu_f,\mu_h)+  \ln U_{S} (\mu_f,\mu_{cs};\nu) + \ln U_D (\mu_f,\mu_c;\nu') \nnb \\
\label{Uevo} 
&&\hspace{1.5cm}= 2 S_{\Gamma} (\mu_h,\mu_{cs})  + \ln\frac{\mu_h^2}{E_J^2 R^2} \cdot a_{\Gamma}(\mu_h,\mu_{cs}) 
- \ln\frac{(\nu'/2)^2}{E_J^2} \cdot a_{\Gamma}(\mu_c,\mu_{cs})  \\
&&\hspace{2cm}+\ln\frac{\nu'^2}{\nu^2} \cdot a_{\Gamma}(\mu_f,\mu_{cs}) -\frac{3C_F}{\beta_0} \ln \frac{\as (\mu_h)}{\as(\mu_c)}\ ,
\nnb  
\eea  
where $U_{H,S,D}$ are the evolution results from the factorization scale to the characteristic scales for $H_J$,~$\tilde{S}_R$, and $\tilde{D}_{\mQ/\mQ}$, respectively. $S_{\Gamma}$ is the Sudakov factor which contains the double logarithmic contributions, 
\be
S_{\Gamma} (\mu_1,\mu_2) = \int^{\mu_1}_{\mu_2} \frac{d\mu}{\mu} \Gamma_C (\as) \ln \frac{\mu}{\mu_1}\ . 
\ee
With the ordinary renormalization scales fixed, the  evolution over $\nu$ is given by
\bea 
&&\ln V(\nu_f,\nu_c,\nu_{cs};\mu,\mu') = \ln V_{S} (\nu_f,\nu_s;\mu) + \ln V_{D} (\nu_f,\nu_c;\mu') \nnb \\
\label{Vevo} 
&&\hspace{2cm}=2 \ln \frac{\nu_s}{\nu_c}\cdot a_{\Gamma}(\mu',1/\bar{b}) + 2 \ln \frac{\nu_f}{\nu_s}\cdot a_{\Gamma}(\mu',\mu)\ .
\eea 

Using the results of $\mu$- and $\nu$-evolutions in Eqs.~\eqref{Uevo} and \eqref{Vevo}, 
we finally obtain the resummed result of $\Phi_{\mQ/J_{\mQ}}$, 
\bea 
\label{TMDJFFres} 
&&\Phi_{\mQ/J_{\mQ}} (z_H, \qp= -\blpu{p}_H/z_H ;E_J,R,m) = H_J (E_J R,\mu_h) \int \frac{db}{2\pi} b J_0 (b|\qp|)  \\ 
&&\hspace{1cm} \times \exp[\mc{M} (\mu_h,\mu_c,\mu_{cs},\nu_c,\nu_{cs}; E_J, R, m, b) ]  
\tilde{S}_R (\bl{b};\mu_{cs},\nu_{cs}) \tilde{D}_{\mQ/\mQ} (z_H,\bl{b};m,\mu_c,\nu_c), \nnb
\eea 
where $\mu_{h,c,cs}$ and $\nu_{c,cs}$ are the characteristic scales for the factorized functions, which, to minimize the large logarithms in the functions, are of the scale
\bea 
\mu_h &\sim& E_JR,~~\mu_c \sim \mu_{cs} \sim q_{\perp} \sim 1/\bar{b}, \label{eq:mu-scales} \\
\nu_c &\sim& 2 E_J,~~\nu_{cs} \sim 2 q_{\perp}/R. 
\eea 
Note that the resummed result in Eq.~\eqref{TMDJFFres} is independent of the factorization scales $\mu_f$ and $\nu_f$. 

The exponentiation factor in Eq.~\eqref{TMDJFFres} is obtained from the suitable combination of Eqs.~\eqref{Uevo} and \eqref{Vevo}, 
\bea 
\mc{M}_R &=& \ln [ U(\mu_f,\mu_h,\mu_c,\mu_{cs};\nu_{cs},\nu_c) \cdot V(\nu_f,\nu_c,\nu_{cs};\mu_f,\mu_f) ] \nnb \\
&=& \ln [ V(\nu_f,\nu_c,\nu_{cs};\mu_{cs},\mu_c) \cdot U(\mu_f,\mu_h,\mu_c,\mu_{cs};\nu_{f},\nu_f) ] \nnb \\
\label{JFFexp} 
&=& 2 S_{\Gamma} (\mu_h,\mu_{cs}) + \ln \frac{\mu_h^2}{E_J^2 R^2} \cdot a_{\Gamma}(\mu_h,\mu_{cs}) + \ln\frac{\nu_c^2}{\nu_{cs}^2}\cdot 
a_{\Gamma}(1/\bar{b},\mu_{cs})  \\
&& - \frac{3 C_F}{\beta_0} \ln \frac{\as(\mu_h)}{\as(\mu_c)} - \ln \frac{\nu_c^2}{4E_J^2} \cdot a_{\Gamma}(\mu_c,\mu_{cs}). \nnb
\eea 
Here we have considered two different evolution paths over $(\mu,\nu)$-plane. In the first line, we first consider the evolution over $\nu$ at $\mu = \mu_f$ and then do the evolution over $\mu$.  In the second line of Eq.~\eqref{JFFexp}, after evolution over $\mu$ with $\nu = \nu_f$, we have performed $\nu$-evolution. Both the evolution results should be the same due to the independence of $\mu$ and $\nu$ scales.  
When we denote a large logarithm as $L$ and power counting it as $\mc{O}(1/\as)$, the first term in the final result of Eq.~\eqref{JFFexp} is dominant and is counted as $\as L^2 \sim \mc{O}(1/\as)$. The next three terms have a size $\as L \sim \mO(1)$. The last term in Eq.~\eqref{JFFexp} is power-counted as $\mc{O}(\as)$, hence it can be ignored at NLL accuracy keeping the large logarithms to $\mO(1)$. 

As studied in subsection~\ref{factmod}, for $q_{\perp} \ll m$ the TMD module has support  in the large $z_H$ region and its factorization is given by Eq.~\eqref{JFFfacbh}. 
Accordingly, the resummed result of $\Phi_{\mQ/J_{\mQ}}$ is 
\bea 
\label{TMDJFFresend} 
&&\Phi_{\mQ/J_{\mQ}} (z_H\to 1, \qp\ll m ;E_J,R,m) = H_J (E_J R,\mu_h) C_{\mQ} (m, \mu_c)\int \frac{db}{2\pi} b J_0 (b|\qp|)  \\ 
&&\hspace{1cm} \times \exp[\mc{M}' (\mu_h,\mu_c,\mu_r,\mu_{cs},\nu_r,\nu_{cs}; E_J, R, m, b) ]  
\tilde{S}_R (\bl{b},\mu_{cs},\nu_{cs}) \tilde{S}_{\mQ/\mQ} (z_H,\bl{b},m,\mu_r,\nu_r), \nnb
\eea 
where $\mu_c \sim m$, and the characteristic scales for $\tilde{S}_R$ and $\tilde{S}_{\mQ/\mQ}$ are given by 
\bea 
\mu_r &\sim& \mu_{cs} \sim q_{\perp} \sim 1/\bar{b} \ll m,  \\
\label{rapisca} 
\nu_r &\sim& 2 E_J(1-z_H)\sim 2E_J \frac{q_{\perp}}{m},~~\nu_{cs} \sim 2 q_{\perp}/R. 
\eea 
Here $\nu_r \gg \nu_{cs}$ since we have the hierarchy $E_JR \gg m$. It is therefore necessary to resum the large logarithms for these very different rapidity scales.  

In Eq.~\eqref{TMDJFFresend}, as a result of the resummation of all the large logarithms to NLL accuracy, the exponentiation factor $\mc{M}'$ is 
\bea 
\mc{M}'_R &=& 2 S_{\Gamma}(\mu_h,\mu_{cs}) - 2 S_{\Gamma}(\mu_c,\mu_{r}) +\ln \frac{\mu_h^2}{E_J^2 R^2} \cdot a_{\Gamma}(\mu_h,\mu_{cs})
- \ln \frac{\mu_c^2}{m^2} \cdot a_{\Gamma}(\mu_c,\mu_{r})  \nnb \\
\label{expfMp}
&&\hspace{-.8cm}-\ln\frac{\nu_r^2}{4E_J^2}\cdot a_{\Gamma}(\mu_r,\mu_{cs}) 
+ \ln\frac{\nu_r^2}{\nu_{cs}^2}\cdot a_{\Gamma}(1/\bar{b},\mu_{cs})
- \frac{C_F}{\beta_0} \Bigl(\ln \frac{\as(\mu_h)}{\as(\mu_{c})}+2\ln \frac{\as(\mu_h)}{\as(\mu_{r})}\Bigr). 
\eea 
Here, the two $S_{\Gamma}$'s are leading terms counted as $\as L^2 \sim \mO(1/\as)$, while the remaining terms are power-counted as $\as L \sim \mO(1)$. 

\section{Heavy hadron's TMD distribution with the thrust axis in $e^+e^-$-annihilation} 
\label{thrust} 

Another interesting application is the heavy hadron's TMD distribution against the thrust axis in $e^+e^-$-annihilation. The TMD distribution for a light hadron has been studied several times in the literature. So it will be interesting to compare those results with the the analysis here when including the heavy quark mass.

\subsection{Resummed results for the heavy hadron's small TMD distribution against the thrust axis} 
\label{thrustres} 

We will consider the small TMD distribution of the heavy hadron that moves into the right hemisphere. 
The situation is very similar to the JFF module in Sec.~\ref{HQTMDJFF}, with the difference here that we consider the hemisphere jet instead of a jet with small radius $R$. Thus, changing the jet size, we can obtain a similar factorization as with the case of the TMD JFF module. 
For simplicity, we consider the production of the heavy quark pair in the dijet limit excluding three jet events in $e^+e^-$-annihilation.

As a result, the double differential cross section for the heavy hadron production with the thrust axis can be factorized as 
\begin{align} 
\frac{1}{\sigma_0} \frac{d\sigma}{dz_H d^2\pp^H} &= \frac{1}{\sigma_0} \frac{d\sigma}{z_H^2 dz_H d^2\qp}  
\nnb \\
&=\frac{2H_{\rm rt}(Q,\mu)}{z_H^2}   \int d^2\kp d^2 \blp{l} S_{\rm rt} (\blp{l},\mu,\nu) D_{H/\mQ} (z_H,\kp,\mu,\nu) \delta^{(2)}(\kp+\blp{l}-\qp) \nnb \\
\label{factthr}
&=\frac{2H_{\rm rt}(Q,\mu)}{z_H^2} \int \frac{db}{2\pi} b J_0 \bigl(\frac{b p_{\perp}^H}{z_H} \bigr) 
\tilde{S}_{\rm rt} (\bl{b},\mu,\nu) \tilde{D}_{H/\mQ}(z_H,\bl{b},\mu,\nu) , 
\end{align}  
where $Q = p^0_{\rm tot}$ is the center of mass energy for the electron and positron, the heavy hadron's energy fraction $z_H = 2p_H\cdot p_{\rm tot}/Q^2 \sim p_H^+/Q$, and $\pp^H$ is the hadron's transverse momentum relative to the thrust axis. We do not distinguish whether the observed hadron from the heavy quark pair production involves the quark or the antiquark, thus the factor of two above.  
$\qp$ is the transverse momentum of the right hemisphere jet (for which the full jet momentum is parallel with the thrust axis) in the hadron frame.  

In Eq.~\eqref{factthr}, $H_{\rm rt}$ is the hard function that contains the hard virtual contributions and radiations in the left hemisphere in the dijet limit. To NLO, it is 
\be
\label{Hrtnlo}
H_{\rm rt} (Q,\mu) = 1+ \frac{\as C_F}{2\pi} \Bigl(-\frac{3}{2} \ln \frac{\mu^2}{Q^2} -\frac{1}{2} \ln^2 \frac{\mu^2}{Q^2} -\frac{9}{2} + \frac{3\pi^2}{4} \Bigr)\ . 
\ee
$S_{\rm rt}$ and its Fourier transform $\tilde{S}_{\rm rt}$ are the soft functions responsible for the soft gluon radiations in the right hemisphere, with the NLO result of $\tilde{S}_{\rm rt}$ being
\be 
\label{Srtnlo}
\tilde{S}_{\rm rt} (\bl{b},\mu,\nu) = 1+\frac{\as C_F}{2\pi} \Bigl(-\ln \bar{b}^2 \mu^2 \ln \frac{\nu^2}{\mu^2} - \frac{1}{2} \ln^2 \bar{b}^2 \mu^2 - \frac{\pi^2}{12} \Bigr)\ . 
\ee

Interestingly, this result can be directly acquired from the result of $\tilde{S}_R$ in Eq.~\eqref{SRnlob} by putting $R \to 2$. In calculating $S_R$ or $\tilde{S}_R$, taking the small $R$ limit, we made the small angle approximation, $\sin R/2 \approx R/2$. 
In the case of the hemisphere soft function, this term becomes $1~(=\sin R/2)$ since $R$ is equal to $\pi$ for this case. 
Thus, with replacement of $R \to 2$, we easily reproduce the result of Eq.~\eqref{Srtnlo}. Similarly,  we can infer the logarithmic terms in $H_{\rm rt}$ from the result of $H_J$ in Eq.~\eqref{HJnlo}. With $R\to 2$, the jet size changes as $E_JR \to 2 E_J \sim Q$, hence the logarithm $\ln \mu/(E_JR)$ in $H_J$ becomes $\ln \mu/Q$ in $H_{\rm rt}$. 

This observation also enables us to resum large logarithms in Eq.~\eqref{factthr} in a remarkably simple way using the result of the TMD JFF module in Sec.~\ref{HQTMDJFF}. The resummed result of Eq.~\eqref{factthr} is
\begin{align} 
\label{formresthr}
\frac{1}{\sigma_0} \frac{d\sigma}{dz_H d^2\pp^H} &=
\frac{2H_{\rm rt}(Q,\mu_h)}{z_H^2} \int \frac{db}{2\pi} b J_0 \bigl(\frac{b p_{\perp}^H}{z_H} \bigr) 
\exp[\mc{M} (\mu_h,\mu_c,\mu_{s},\nu_c,\nu_{s}; Q, \bar{b}) ]  \nnb \\
&\hspace{3cm}\times 
\tilde{S}_{\rm rt} (\bl{b},\mu_s,\nu_s) \tilde{D}_{H/\mQ}(z_H,\bl{b},\mu_c,\nu_c),  
\end{align} 
where the exponentiation factor $\mc{M}$ can be directly obtained from the result of Eq.~\eqref{JFFexp} setting $R\to 2$. It reads 
\begin{align} 
\mc{M}_T (\mu_h,\mu_c,\mu_{s},\nu_c,\nu_{s}; Q, \bar{b}) &= 2 S_{\Gamma} (\mu_h,\mu_{s}) + \ln \frac{\mu_h^2}{Q^2} \cdot a_{\Gamma}(\mu_h,\mu_{s}) + \ln\frac{\nu_c^2}{\nu_{s}^2}\cdot 
a_{\Gamma}(1/\bar{b},\mu_{s})  \nnb \\
\label{expfthr}
& - \frac{3 C_F}{\beta_0} \ln \frac{\as(\mu_h)}{\as(\mu_{c})} - \ln \frac{\nu_c^2}{Q^2} \cdot a_{\Gamma}(\mu_c,\mu_{s}). 
\end{align} 
Here the characteristic scales are given by 
\be
\mu_h \sim \nu_c\sim Q,~~\mu_c \sim \mu_{s} \sim q_{\perp} \sim 1/\bar{b}. 
%\nu_c &\sim 2 E_J,~~\nu_{cs} \sim 2 q_{\perp}/R. 
\ee

If $b \gg 1/m$, similar to the heavy quark TMDFF, we can use the refactorization results in Eq.~\eqref{HQFFF}, hence the resummed result is
\begin{align} 
\label{formresthrend}
\frac{1}{\sigma_0} \frac{d\sigma}{dz_H d^2\pp^H} &=
\frac{2H_{\rm rt}(Q,\mu_h)}{z_H^2} \int \frac{db}{2\pi} b J_0 \bigl(\frac{b p_{\perp}^H}{z_H} \bigr) 
\exp[\mc{M}' (\mu_h,\mu_c,\mu_r,\mu_{s},\nu_r,\nu_{s}; Q, \bar{b}) ]  \nnb \\
&\hspace{2.5cm}\times 
C_{\mQ}(m,\mu_c)\tilde{S}_{\rm rt} (\bl{b},\mu_s,\nu_s) \tilde{S}_{H/\mQ}(z_H,\bl{b},\mu_r,\nu_r).   
\end{align} 
From the result of Eq.~\eqref{expfMp} with $R\to 2$, we obtain 
\begin{align} 
\mc{M}'_T(\mu_h,\mu_c,\mu_r,\mu_{s},\nu_r,\nu_{s}; Q, \bar{b})
&= 2 S_{\Gamma}(\mu_h,\mu_{s}) - 2 S_{\Gamma}(\mu_c,\mu_{r}) +\ln \frac{\mu_h^2}{Q^2} \cdot a_{\Gamma}(\mu_h,\mu_{s})
 \nnb \\
&\hspace{-.8cm}- \ln \frac{\mu_c^2}{m^2} \cdot a_{\Gamma}(\mu_c,\mu_{r}) 
-\ln\frac{\nu_r^2}{Q^2}\cdot a_{\Gamma}(\mu_r,\mu_{s}) 
+ \ln\frac{\nu_r^2}{\nu_{s}^2}\cdot a_{\Gamma}(1/\bar{b},\mu_{s}) \nnb \\
\label{expfMpthr}
&\hspace{-.8cm} -\frac{C_F}{\beta_0} \Bigl(\ln \frac{\as(\mu_h)}{\as(\mu_{c})}+2\ln \frac{\as(\mu_h)}{\as(\mu_{r})}\Bigr)\ ,  
\end{align}
where the characteristic scales are estimated to be
\begin{align}
\mu_h &\sim Q,~\mu_c \sim m,~\mu_r \sim \mu_{s} \sim  1/\bar{b},  \\
\label{rapiscathr} 
\nu_r &\sim Q (1-z_H)\sim \frac{Q}{m\bar{b}},~~\nu_{s} \sim 1/\bar{b}. 
\end{align} 

\subsection{Numerical analysis for the resummed result} 

In this subsection, we show numerical results for the TMD distribution with respect to the thrust axis combining the resummed results of Eqs.~\eqref{formresthr} and \eqref{formresthrend} in the subsection~\ref{thrustres}. Here we focus on the region where $p_T^H$ is small, but mostly perturbative, i.e, $\Lambda_{\rm{QCD}} \lesssim p_T^H \lesssim m$, hence the perturbative TMDFF can be matched onto the nonperturbative FF, $\phi_{H/\mQ}(z)$, as illustrated in Eq.~\eqref{npimp} (also Eq.~\eqref{tDbs} or Eq.~\eqref{tDbl} in $\bb$-space).  
 
As a result, we provide the formalism for the numerical implementation of the resummed results,
\begin{align} 
\frac{1}{\sigma_0} \frac{d\sigma}{dz_H dp_{\perp}^H} &=
\frac{2 p_{\perp}^HH_{\rm rt}(Q,\mu_h)}{z_H^2} \int  db b J_0 \bigl(\frac{b p_{\perp}^H}{z_H} \bigr) 
\exp[\mc{M}_{\mr{NP}}] \cdot \tilde{S}_{\rm rt} (b^*,\mu_s,\nu_s)  \nnb \\
&\times 
\int^1_{z_H} \frac{dz}{z}  \phi_{H/\mQ} \bigl(\frac{z_H}{z}\bigr) \Biggl[\exp[\mc{M}^L_{\mr{P}}]\cdot
C_{\mQ}(m,\mu_c)\tilde{S}_{\mQ/\mQ}(z,b^*,\mu_r,\nu_r) \label{numform}\\
&\hspace{2cm}+\exp[\mc{M}_{\mr{P}}^F]\cdot \Delta(z,b^*,\mu_c) \Biggr]\ .\nnb
\end{align} 
Here $p_{\perp}^H = |\pp^H|$, and in order to avoid the Landau pole as $b$ becomes large we have expressed the cross section in $\bb$-space using $b^*$ rather than $b$. Following the prescription introduced in Ref.~\cite{Collins:1984kg}, $b^*$ has been given by   
\be 
b^* = \frac{b}{\sqrt{1+b^2/b_{\rm{max}}^2}}\ .
\label{eq:bstar}
\ee
So, in the perturbative region where $b$ is small ($b \ll b_{\rm{max}}$), $b^*$ is given to be $b^* \approx b$. But, when $b$ becomes large, $b^*$ becomes frozen at $b_{\rm{max}}$. Here our default choice of $b_{\rm{max}}$ will be $b_{\rm{max}}=2~\rm{GeV}^{-1}$ in order for checking perturbative effects maximally. With the choice, the freezing scale for $\as$ is given by $\mu_{fr} \sim 1/(b_{\rm{max}} e^{\gamma_E}/2) \sim 0.56~\rm GeV$. 

In Eq.~\eqref{numform}, the perturbative exponential factors $\mc{M}_P^L$ and $\mc{M}_P^F$ respectively represent 
$\mc{M}'_T$ in Eq.~\eqref{expfMpthr} and $\mc{M}_T$ in Eq.~\eqref{expfthr} with replacement of $b \to b^*$. We define $\Delta(z,b^*,\mu_c)$ as the difference between the perturbative TMDFFs for $b \sim 1/m$ and $b \gg 1/m$,  given by  
\begin{align} 
\Delta(z,b^*,\mu_c) = \tilde{D}_{\mQ/\mQ} (z,b^*,\mu_c,\nu) - 
C_{\mQ} (m,\mu_c)  \tilde{S}_{\mQ/\mQ} (z,b^*,\mu_c,\nu). 
\end{align}
$\Delta(z,b^*,\mu_c)$ does not include  large logarithms of $1-z$ and $b$, and is independent of the rapidity scale $\nu$. At order $\as$, it is 
\begin{align} 
\Delta(z,b,\mu) =& \frac{\as(\mu) C_F}{2\pi} \Bigl\{ \frac{4}{1-z} \Bigl[z K_0\bigl(\frac{1-z}{z} mb \bigr)
-  K_0\bigl((1-z) mb\bigr)\Bigr] + 2(1-z) K_0 \bigl(\frac{1-z}{z} mb \bigr) \nnb \\
\label{deltaNLO}
&~~~-2 b m \Bigl[K_1\bigl(\frac{1-z}{z} mb \bigr)
-  K_1\bigl((1-z) mb\bigr)\Bigr] \Bigr\} \ . 
\end{align}

For the full description to the entire large $b$ (or small $p_T^H$) region, we have also introduced the nonperturbative factor $\mc{M}_{\rm NP}$ in Eq.~\eqref{numform}. 
Basically, it is introduced to parameterize  hadronization effects and in principle could be obtained from the fit to experiment data as done in case of TMDFF to a light meson~\cite{Kang:2017glf,Kang:2020yqw,Sun:2014dqm,Kang:2015msa}.
Due to a lack of experimental data on the TMD distribution of the heavy meson, we do not try to extract a specific parameterization of $\mc{M}_{\rm NP}$ nor try to do a more sophisticated approach, e.g., like a recent analysis that separates short and long distance contributions to TMD distribution for a light quark~\cite{Ebert:2022cku}. These are beyond the scope of this paper. Instead we introduce a simple model: 
\be 
\mc{M}_{\rm NP} = (-1)^{1+p} \Bigl(1-\frac{b}{b^*}\Bigr)^p, 
\ee
where $p$ is a positive integer, and our default choice will be $p=2$. 
The role of $\mc{M}_{\rm NP}$ here is to monotonously connect the perturbative result to the nonperturbative  $p_T$ region.

\begin{figure}[h]
	\begin{center}
		\includegraphics[width=15.5cm]{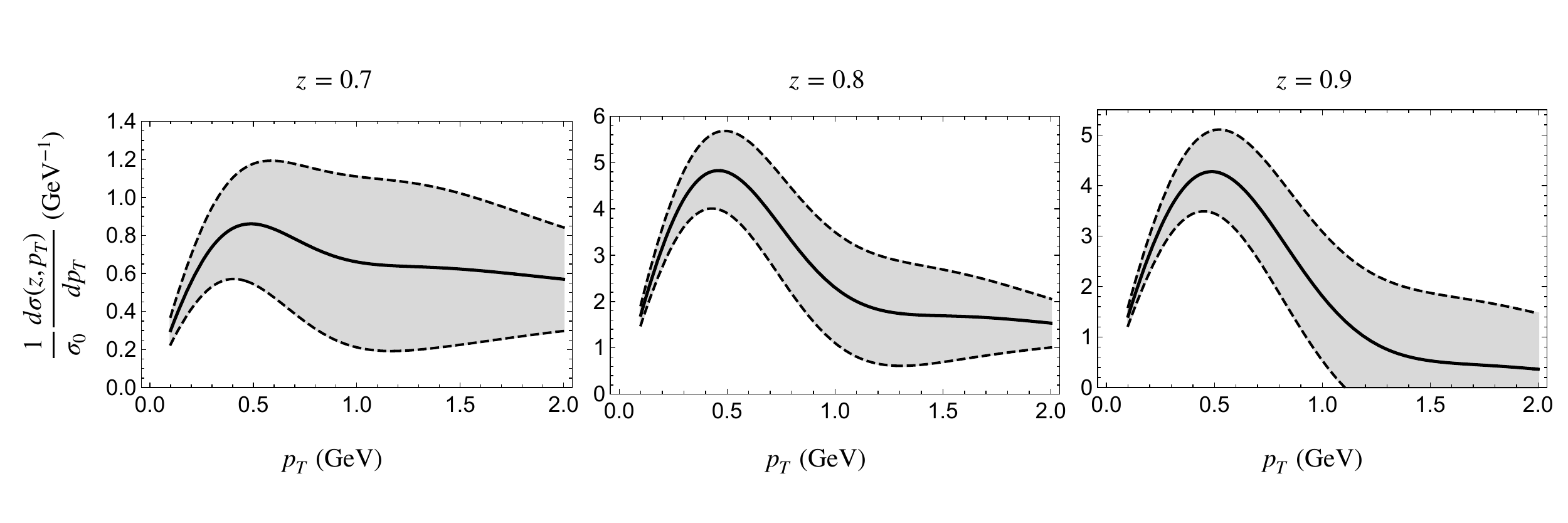}
	\end{center}
	\vspace{-1cm}
	\caption{\label{fig:pt} TMD distribution for a possible $b$-flavored heavy-light hadron inside a hemisphere jet in $e^+e^-$ collisions. For simplicity, we have short-written $z_H$ and $p_T^H$  as $z$ and $p_T$ respectively. The center of mass energy of the collision $Q = 100 ~ \rm{GeV}$ and $b_{\rm max} = 2 ~\mathrm{GeV}^{-1}$.}
\end{figure}

In Fig.~\ref{fig:pt} we show TMD distributions of single $b$-flavored heavy-light hadron inside a hemisphere jet in  $e^+e^-$ collisions (and not specified the $b$-hadron so $N_H$ in Eq.~\eqref{npFF} is set to 1), with energy fraction $z_H$ carried by the heavy hadron fixed. The error bands come from varying each characteristic scale $\mu_i (\nu_i)$ that appears in Eq.~\eqref{numform} up to $2\mu_i (2\nu_i)$ and down to $\mu_i/2 (\nu_i/2)$, and summing the errors from all the scale variations by quadrature. As $p_T^H$ approaches $0$, the error bands get narrower. This is because small $p_T$ lies in the non-perturbative region, and we simply freeze out the scale variations in those regions. That is, we only estimate the error from our perturbative computations, since we do not have control of the error of non-perturbative origin. We put more details on how we treat the scale variations involving non-perturbative regions in Appendix~\ref{sec:scale-var}. 

\begin{figure}[h]
	\begin{center}
		\includegraphics[width=10cm]{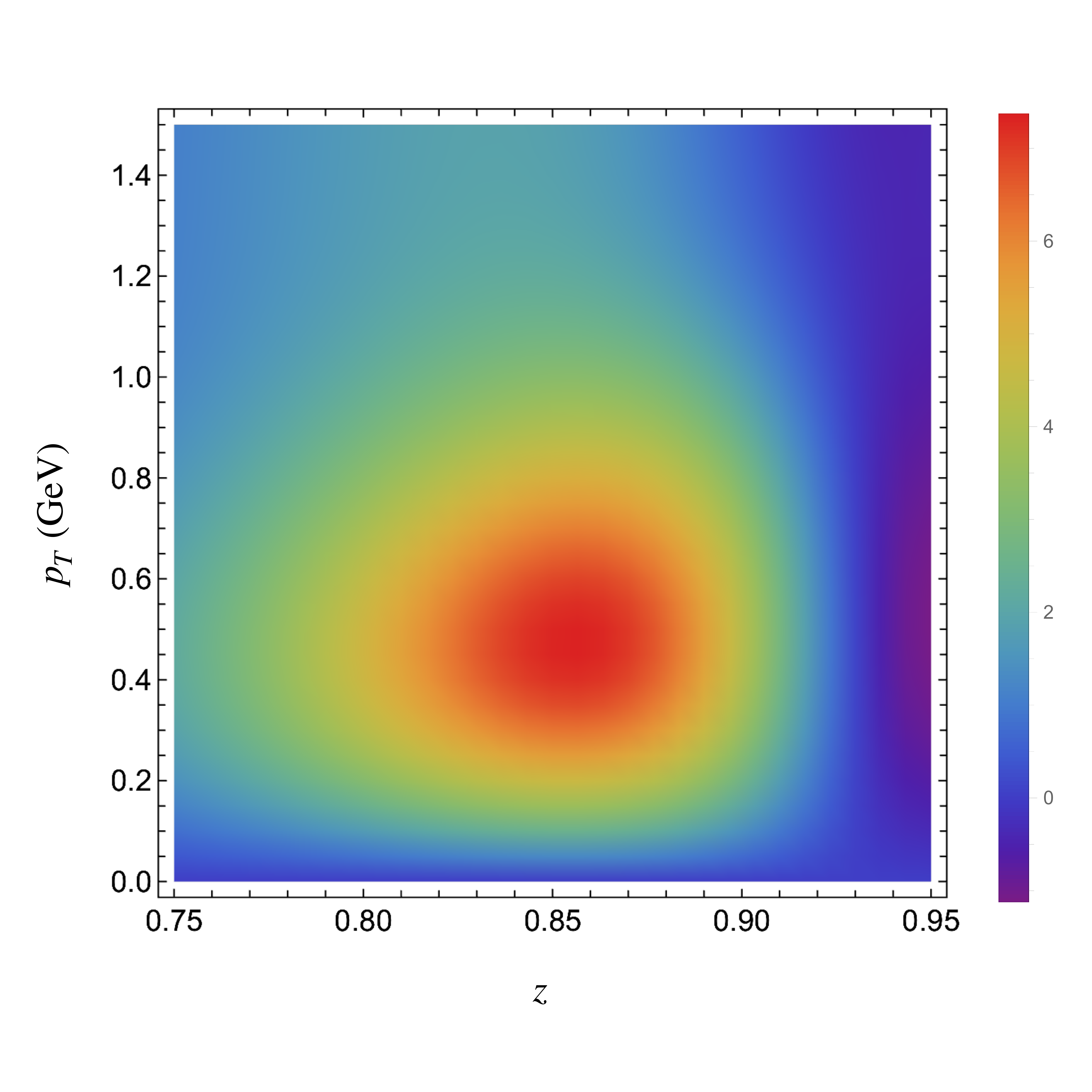}
	\end{center}
	\vspace{-1cm}
	\caption{\label{fig:zp} Two dimensional contour plot of $(z,p_T)$ distribution for a b-flavored heavy-light hadron in a hemisphere jet in $e^+e^-$ collisions, according to the cross section Eq.~\eqref{numform} for numerical evaluations.} 
\end{figure}

To have a better view of the joint $(z_H,p_T^H)$ distributions as displayed in Eq.~\eqref{numform}, we made a two-dimensional contour plot  shown in Fig.~\ref{fig:zp}, where all the parameters are the same as those used in Fig.~\ref{fig:pt}, using the central value of the scales.
As shown in the contour plot, the dominant contributions come from the range roughly $z_H \in [0.8,0.9]$ and $p_T^H \in [0.3, 0.7]~\rm GeV$. Even though the peak region is close to the nonperturbative domain, hence we need more sophisticated parameterization and study of the hadronization, we suspect that the shape of the distribution in Fig.~\ref{fig:zp} show some characteristics for a heavy-light hadron with a $b$ quark.   
Note that in Fig.~\ref{fig:zp}, some negative values appear near the right edge of the plot, which needs more clarified studies on hadronization effects because they too close to the non-perturbative region. For instance, $z = 0.95$ means that the residual scale for a $B$ meson $(1-z)m_B$ is around $0.25$~GeV.

\section{Conclusions}
\label{conclusions}

In this paper, we study the heavy quark (HQ) mass effects to the transverse momentum dependent fragmentation function (TMDFF) using SCET. We start by calculating the one-loop contribution to the TMDFF that is initiated by a heavy quark.  The resulting function is IR finite. While the IR dependence of the HQ TMDFF is different than the light quark case, the UV divergence comes from the virtual contribution, and thus is the same as found in the light TMDFF.  The rapidity divergence comes from the zero-bin subtraction, and thus is also the same as the light TMDFF. 

Given the possible hierarchy  of scales between $q_\perp$ and $m$, where $q_\perp$ is the transverse momentum of the initiating parton with respect to hadron and $m$ is the heavy quark mass, we investigate the HQ TMDFF in the limit  $q_\perp \ll m$. This is done by matching onto boosted heavy quark effective theory. This allows us to factorize the HQ TMDFF further into a shape function and a matching coefficient, which is done at one-loop order.  We next study the opposite limit, $q_\perp \gg m$. In this case, we integrate out the fluctuations of $q_\perp$ and match onto the standard heavy quark fragmentation function. This is again done at one-loop. Finally, since the nonperturbative effects are always important when describing the hadronization of the final state hadron, we also include the nonperturbative fragmentation function, using a model previously introduced in the literature. 

Using the above results, we study two different applications. First we study the heavy quark TMD jet fragmentation function (JFF), which describes a heavy quark fragmenting to a jet, where inside the jet is an observed heavy hadron. By studying this process, we may gain useful information of the hadronization of the heavy quark. When $q_\perp$ is much smaller than the jet scale, we can further factorize the HQ TMD JFF into the standard FF and what we define as the JFF module, containing the transverse momentum dependence. The JFF module can be factored into a hard function, a soft function, and the HQ TMDFF. We resum leading large logarithms (not including nonglobal logarithms) in the JFF module to NNL order.

As a second application, we investigate the heavy hadron TMD distribution with respect to the thrust axis in $e^+e^-$ annihilation. The results can be resummed using the  HQ TMD JFF we obtained and numerical results are shown. In order to produce sensible results, we have a better handle on the nonperturbative region, but a more in depth study is beyond the scope of this paper.

\acknowledgments

LD is supported by the Alexander von Humboldt Foundation. CK is supported by Basic Science Research Program through the National Research Foundation of Korea (NRF) funded by the Ministry of Science and ICT (Grant No. NRF-2021R1A2C1008906). AKL is supported in part by the National Science Foundation under Grant No. PHY-2112829.

%%%%%%%%%%%%%%%%%%%%%%%%%%%%%%%%%%%%%%%%%%%%%%%%%%%%%%%%%%%%%%%%%%%%%%
%%%%%%%%%%%%%%%%%%%%%%%%%%%%% Appendix %%%%%%%%%%%%%%%%%%%%%%%%%%%%%%%
%%%%%%%%%%%%%%%%%%%%%%%%%%%%%%%%%%%%%%%%%%%%%%%%%%%%%%%%%%%%%%%%%%%%%%

\appendix

\section{NLO result of the heavy quark TMDFF at parton frame}
\label{FFp}

In the parton frame where the initiating parton is taken to have zero transverse momentum, the heavy quark TMDFF in $D$ dimension is 
given by
\be 
\label{defTMDFFp} 
\mD_{H/\mc{Q}} (z,\blp{p},\mu,\nu) = \sum_X \frac{1}{2N_c z} \mr{Tr} \langle 0 | \delta \left(\frac{p_+}{z} - \mc{P}_+ \right) \delta^{(D-2)} (\bsp{\mP}) \nn \Psi_n^{\mQ} | H(p) X \rangle  \langle H(p) X | \bar{\Psi}_n^{\mQ} | 0 \rangle . 
\ee
Here the derivative operator returns the transverse momentum of the initial parton 
expressed as $\bsp{\mP} =\pp +  \blpu{p}_X = 0$. In this case the fragmentation function is the distribution of the transverse momentum for the observed hadron, $\pp$, which, as 
introduced in Eq.~\eqref{relfra}, is related to the transverse momentum of the initial parton in the hadron frame by 
$\qp = -\pp/z$.  
So we have the following explicit relation between the fragmentation functions in the parton and at the hadron frames:\footnote{
Note that Eq.~\eqref{fragrel} is no more than the probability density for finding a hadron with a large momentum fraction $z$ and a small transverse momentum $\pp$~\cite{Collins:1981uw}. 
}
\be 
\label{fragrel}
\mD_{H/f} (z,\pp,\mu,\nu)  =  D_{H/f} (z,-\pp/z,\mu,\nu).
\ee
This relation holds for any flavor of parton $f$. 

Similar to how we obtained the NLO result of the fragmentation function at hadron frame, we can compute the NLO correction to the fragmentation function in the parton frame. The bare one-loop result in  momentum space is 
\bea 
\label{DQQmnlop}
\mD_{\mQ/\mQ}^{(1)} (z,\blp{p},\mu,\nu) 
&=&\frac{\as C_F}{2\pi^2} \Biggl\{ \delta(1-z) \delta(\pps) 
\Biggl[\Bigl(\frac{2}{\eta} + 2\ln\frac{\nu}{p_+} +\frac{3}{2} \Bigr) \Bigl(\frac{1}{\euv} + \ln\frac{\mu^2}{\Lambda^2} \Bigr)+2 \\ 
&& -\ln(1+\lambda) - \frac{2}{\sqrt{\lambda}} \arctan{\sqrt{\lambda}} - \mr{Li}_2 (-\lambda) \Biggr] 
-\delta(\pps) \Biggl[\frac{P_{qq}(z)}{C_F} \ln\lambda \nnb \\
&&+\left(\frac{2z}{1-z} \Bigl(\ln\frac{1+(1-z)^2 \lambda}{(1-z)^2} - \frac{1}{1+(1-z)^2 \lambda}\Bigr)\right)_+
+ (1-z) \ln\frac{1+(1-z)^2 \lambda}{(1-z)^2}\Biggr] \nnb \\ 
&&-\Bigl(\frac{2}{\eta} + 2\ln\frac{\nu}{p_+} +\frac{3}{2} \Bigr) \delta(1-z) \left(\frac{1}{\pps}\right)_{\Lambda^2} + \frac{P_{qq}(z)}{C_F} \left(\frac{1}{\pps+(1-z)^2 m^2}\right)_{\Lambda^2}
\nnb \\
&&-2z(1-z) \left(\frac{m^2}{(\pps + (1-z)^2 m^2)^2} \right)_{\Lambda^2} \Biggr\}. \nnb 
\eea 
Here the rapidity and UV divergences are the same as for the fragmentation function at hadron frame. 

In impact-parameter space, the renormalized one-loop result is 
\bea 
\tilde{\mD}_{\mQ/\mQ} (z,\bb;\mu,\nu) 
&=& \int d^2\blp{p} e^{i\bl{b} \cdot \blp{p}} \mD_{\mQ/\mQ} (z,\blp{p};\mu,\nu) \nnb \\
&=& 1+ \frac{\as C_F}{2\pi}  \Biggl\{ \delta(1-z) \Bigl[\Bigl(2\ln\frac{\nu}{p_+}+\frac{3}{2} \Bigr) \ln \bar{b}^2 \mu^2 + \frac{1}{2} \ln \bar{b}^2 m^2 \Bigr] \nnb \\
\label{DQQ1bp} 
&&+\Bigl(\frac{2z}{1-z}\Bigr)_+ \Bigl[2K_0 ((1-z)mb) +2\ln(1-z) -1\Bigr] \\
&&+2(1-z)K_0 ((1-z)mb)-\Bigl(\frac{4z}{1-z} \ln(1-z)\Bigr)_+  \nnb \\
&&-2z(1-z) \Bigl[\frac{bm}{1-z} K_1 ((1-z)mb) -\frac{1}{(1-z)^2} \Bigr]\Biggr\}\ . \nnb
\eea 
Here $b \sim 1/p_{\perp}$. This result in the parton frame can be easily compared with the hadron frame result, Eq.~\eqref{DQQ1b}, where $b \sim 1/q_{\perp} = z/p_{\perp}$. From the result of Eq.~\eqref{DQQ1b} with replacement $b \to z b$, we immediately obtain the result Eq.~\eqref{DQQ1bp}.

We can also consider the heavy quark fragmentation in the parton frame in the limit $\pp \ll m$. In this case, the same factorization as Eq.~\eqref{HQFFF} holds and the fragmentation function is given by 
\be 
\label{HQFFFp} 
\mD_{H/\mc{Q}} (z,\blp{p}\ll m,\mu,\nu) = C_{\mQ} (m,\mu) S_H (z,\blp{p},\mu,\nu). 
\ee
Note that the heavy quark shape function $S_H$ is the same as for the hadron frame, with the one-loop result at the parton level given in Eq.~\eqref{SQnlo}.  As explained in Sec.~\ref{smallq}, the fragmentation for the small $p_{\perp}$ region is actually described by the residual mode in bHQET, which contributes to only for the large $z$ region. 
Thus, at leading power of $1-z$, the transverse momenta for the parton and hadron frames can be identified, 
\be 
|\qp| = \frac{|\pp|}{z} \sim |\pp|. 
\ee

\section{One loop calculation of the TMD csoft function}
\label{TMDcsoft}
In this section we perform the one-loop calculation the TMD csoft function $S_R$ defined in Eq.~\eqref{defSR}, reproduced here for convenience
\be 
\label{defSR2} 
S_R (\lp;\mu,\nu) = \frac{1}{N_c} \mr{Tr}~\langle 0 | \tilde{Y}_{n,cs} Y_{\n,cs}^{\dagger} \delta^{(2)} (\lp + \Theta_{in} \cdot  \bsp{\mP})  Y_{\n,cs} \tilde{Y}_{n,cs}^{\dagger} |0\rangle.
\ee 
As expressed in the argument of the delta function in Eq.~\eqref{defSR2}, the csoft function returns a nonzero value of $\lp$ only when at least one gluon is radiated inside of the jet, while the delta function becomes $\delta^{(2)} (\lp)$ for gluons that are all radiated outside of the jet. 

\begin{figure}[h]
\begin{center}
\includegraphics[height=8cm]{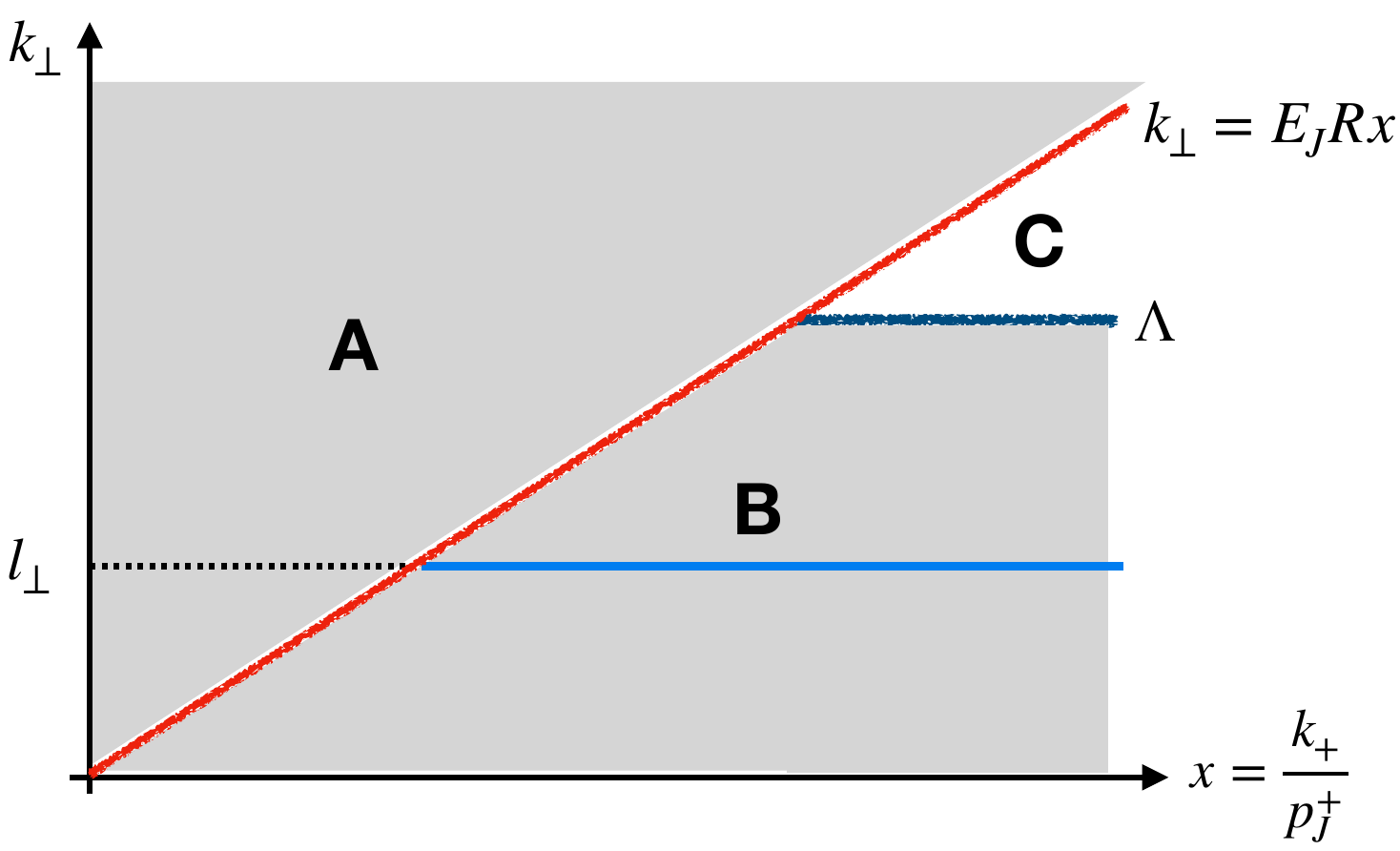}
\end{center}
\vspace{-0.5cm}
\caption{\label{fig4}
The phase space for real gluon emission for the calculation of the TMD csoft function $S_R$. 
The red line denotes the jet boundary, and the blue solid line in the region `{\bf B}' shows the contribution to the distribution with a nonzero $l_{\perp}~(|\equiv\lp|)$. 
}
\end{figure}

In Fig.~\ref{fig4} we have illustrated the phase space for a real gluon emission for the one loop calculation. 
Here the csoft gluon momentum $k^{\mu}$ is power counted as shown in Eq.~\eqref{csoftsca} 
and we consider the limit, $k_{\perp} ~(\equiv |\kp|) \ll E_JR$. 
Hence the largest momentum component $k_+$ should be much smaller than $p_J^+~(\sim 2E_J)$ 
in the power counting, and the jet boundary can be approximated to be $k_{\perp} = E_J R x$, where $x = k_+/p_J^+$. However, when integrating over $k$, the limit for $k_+$ can be set to be infinity, r since the momentum $p_J^+$ is to be considered infinitely larger than the csoft momentum.

When we consider the real gluon emission inside the jet, the transverse momentum is $l_{\perp}~(\equiv |\lp|)$, and the amplitude is given by 
\be 
M_{in}^R (\lps) = \frac{\as C_F}{\pi^2} \frac{(\mu^2 e^{\gamma_E})^{\eps}}{\Gamma(1-\eps)} \left(\frac{\nu}{p_J^+}\right)^{\eta} \Bigl(\frac{1}{\lps}\Bigr)^{1+\eps} \int^{\infty}_{l_{\perp}/E_JR} dx x^{-1-\eta}\ ,  
\ee
where we employed the rapidity regulator in order to handle the divergence as $x\to \infty$.
$M_{in}^R$ has an IR divergence as $\lps \to 0$, hence in order to regulate we use the $\Lambda^2$-distribution, 
\be
\label{MinR}
M_{in}^R (\lps) = \Bigl[\int^{\Lambda^2}_0 d\kps M_{in}^R (\kps) \Bigr] \delta (\lps) + \Bigl[M_{in}^R (\lps) \Bigr]_{\Lambda^2}\ .  
\ee
The integration region of the first term with the delta function covers the region `{\bf B}' in the phase space shown in Fig.~\ref{fig4}.  

The out-jet region for real emission, where the amplitude is proportional to $\delta(\lps)$, coincides with the region `{\bf A}' in Fig.~\ref{fig4}. Therefore, if we combine the virtual contribution and the contributions from the integration of the regions `{\bf A}' and `{\bf B}', the net contribution becomes the result of the integration of the region `{\bf C}' with an overall negative sign since the virtual contribution covers the full phase space of Fig.~\ref{fig4} with the opposite sign.   
Thus, the net contribution proportional to $\delta(\lps)$ is
\begin{align} 
\mc{M}_{\delta} &= -\frac{\as C_F}{\pi^2} \frac{(\mu^2 e^{\gamma_E})^{\eps}}{\Gamma(1-\eps)} \left(\frac{\nu}{p_J^+}\right)^{\eta} \int^{\infty}_{\Lambda/E_JR} dx x^{-1+\eta} \int^{x^2 E_J^2 R^2}_{\Lambda^2} d\kps (\kps)^{-1-\eps} \nnb \\
\label{Mdelta} 
&= \frac{\as C_F}{2\pi^2} \Bigl[\frac{1}{\eps^2} + \frac{1}{\eps} \ln\frac{\mu^2}{\Lambda^2} + \frac{1}{2} \ln^2\frac{\mu^2}{\Lambda^2} - \frac{\pi^2}{12} -2 \Bigl(\frac{1}{\eps} + \ln\frac{\mu^2}{\Lambda^2}\Bigr) 
\Bigl(\frac{1}{\eta} + \ln\frac{\nu R}{2\Lambda}\Bigr) \Bigr]\ , 
\end{align} 
where  the $1/\eps$ poles are due to the UV divergences. 

The remaining contribution for the one-loop calculation of $S_R$ is the second term in Eq.~\eqref{MinR}, i.e., the $\Lambda^2$ distribution of $M_{in}^{R}$ with  nonzero $\lps$, for which the integration region is denoted as the blue solid line in region `{\bf B}' of Fig.~\ref{fig4}. Since $\lps\neq 0$, $M_{in}^R$ is free from the IR divergence and is computed as 
\begin{align} 
M_{in}^R (\lps \neq 0) &= \frac{\as C_F}{\pi^2} \left(\frac{\nu}{p_J^+}\right)^{\eta} 
\frac{1}{\lps}\int^{\infty}_{l_{\perp}/E_JR} dx x^{-1+\eta} \nnb \\
\label{MRdist} 
&= \frac{\as C_F}{\pi^2} \frac{1}{\lps} \Bigl(\frac{1}{\eta} + \frac{1}{2} \ln \frac{\nu^2 R^2}{4\lps} \Bigr)\ . 
\end{align}

Combining the results of Eqs.~\eqref{Mdelta} and \eqref{MRdist}, we obtain the one-loop result of the csoft function $S_R$ as 
\begin{align} 
S_R^{(1)} (\lp,\mu,\nu) &= \frac{\as C_F}{2\pi^2} \Bigl\{\delta(\lps)\Bigl[\frac{1}{\eps^2} + \frac{1}{\eps} \ln\frac{\mu^2}{\Lambda^2} + \frac{1}{2} \ln^2\frac{\mu^2}{\Lambda^2} - \frac{\pi^2}{12} -2 \Bigl(\frac{1}{\eps} + \ln\frac{\mu^2}{\Lambda^2}\Bigr) 
\Bigl(\frac{1}{\eta} + \ln\frac{\nu R}{2\Lambda}\Bigr) \Bigr] \nnb \\
\label{SRonel}
&\hspace{2cm}+\Bigl[\frac{1}{\lps} \Bigl(\frac{2}{\eta} + \ln \frac{\nu^2 R^2}{4\lps} \Bigr)\Bigr]_{\Lambda^2} \Bigr\}\ . 
\end{align} 
The renomalized result and the result in $\bb$-space are presented in Eqs.~\eqref{SRnlom} and \eqref{SRnlob}, respectively. 
Furthermore, as discussed in Sec.~\ref{thrust}, we can obtain the one-loop result of the TMD soft function with thrust axis by setting $R \to 2$.

\section{Implication of nonperturbative contributions for $\qp \sim \Lambda_{\rm QCD}$} 
\label{NPq}

When $\qp \sim \Lambda_{\rm QCD}$, the transverse momentum distribution becomes entirely nonperturbative. 
Since the heavy quark mass is taken to be much larger than $\qp$, we can integrate out the degrees of freedom of the scale $p^2 \sim m^2$ and obtain the heavy quark function $C_{Q}(m,\mu)$ before we consider the nonperturbative TMD function. Therefore the heavy quark TMD FF for $\qp \sim \Lambda_{\rm QCD}$ can be written as 
\be 
\label{HQTMDFFnp} 
D_{H/\mc{Q}} (z,\blp{q} \sim \Lambda_{\rm QCD} ;\mu,\nu) = C_{\mQ} (m,\mu) S_{H/\mQ} (z,\blp{q}\sim \Lambda_{\rm QCD},\mu,\nu),  
\ee
where $S_{H/\mQ}$ has been introduced in Eq.~\eqref{SHdef} and in this case is totally nonperturbative. 

The rapidity scale dependence in $S_{H/\mQ}$ complicates any nonperturbative parameterization and its modeling. 
However, when we consider the whole scattering process, there will be another nonperturbative TMD soft function also with rapidity scale dependence. 
When combined with $S_{H/\mQ}$, as seen in Eq.~\eqref{JFFfacbh}, the rapidity scale dependence can be removed.   
Therefore, for example, when we consider the nonperturbative TMD distribution of the HQTMD JFF studied in Sec.~\ref{HQTMDJFF}, it is useful to introduce a new function combining with $S_R$ in Eq.~\eqref{defSR}: 
\be 
\label{SHQR}
\tilde{\mc{S}}_{H/\mQ}^R (z,\bl{b}; \mu)  = \tilde{S}_R (\bl{b};\mu,\nu) \tilde{S}_{H/\mQ} (z,\bl{b};m,\mu,\nu).
\ee
Although $\tilde{\mc{S}}_{H/\mQ}^R$ is not dependent of the rapidity scale, it involves a large logarithm that comes from the rapidity gap between $\tilde{S}_R$ and $\tilde{S}_{H/Q}$,
%in the right side of eq.~\eqref{SHQR}, 
\be 
\ln \frac{\nu_{cs}}{\nu_r} \approx \ln \frac{2q_{\perp}/R}{2E_J q_{\perp}/m} = \ln \frac{m}{E_JR},  
\ee
where $\nu_s$ and $\nu_{r}$ are the characteristic rapidity scales of $\tilde{S}_R$ and $\tilde{S}_{H/Q}$, respectively,  shown in Eq.~\eqref{rapisca}. 
In the perturbative limit, resumming the large rapidity logarithms gives
\be 
\label{SHQR1} 
\tilde{\mc{S}}_{H/\mQ}^R (z,\bl{b}; \mu)  = \Bigl(\frac{\nu_{cs}}{\nu_r}\Bigr)^{2a_{\Gamma} (\mu, 1/\bar{b})} 
\tilde{S}_R (\bl{b};\mu,\nu_s) \tilde{S}_{H/\mQ} (z,\bl{b};m,\mu,\nu_r).
\ee

The $\mu$-evolution result for the combined function, $\tilde{\mc{S}}_{H/Q}^R$, is given by 
\be 
\tilde{\mc{S}}_{H/\mQ}^R (z,\bl{b}; \mu) = U_{\mc{S}}^R (\mu, \mu_0)  \tilde{\mc{S}}_{H/\mQ}^R (z,\bl{b}; \mu_0),  
\ee
where the evolution kernel at NLL is 
\be 
\ln U^R_{\mc{S}} (\mu, \mu_0) = \ln \frac{m^2}{E_J^2 R^2} a_{\Gamma} (\mu,\mu_0) 
- \frac{2C_F}{\beta_0} \ln \frac{\as (\mu)}{\as (\mu_0)}\ .
\ee
Here, in order to guarantee a perturbative expansion, the lower scale $\mu_0$ must be chosen as some scale  above $\Lambda_{\mr{QCD}}$, e.g., $\mu_0 \sim 1~\mr{GeV}$. Then we can parameterize 
$\tilde{\mc{S}}_{H/\mQ}^R (z,\bl{b}; \mu_0)$ as a genuine nonperturbative function. 

When we consider  heavy hadron fragmentation with respect to the thrust axis, studied in Sec.~\ref{thrust}, following discussions in Refs.~\cite{Collins:2011zzd,Ebert:2019okf,Kang:2020yqw}, 
the nonpertubative TMD function can be defined as 
\be 
\label{SHQrt}
\tilde{\mc{S}}_{H/\mQ}^{rt} (z,\bl{b}, \mu)  = \tilde{S}_{rt} (\bl{b},\mu,\nu) \tilde{S}_{H/\mQ} (z,\bl{b},\mu,\nu).
\ee
Here the large rapidity logarithms are induced from the large gap between the characteristic scales $\nu_r$ and $\nu_s$, given by 
\be 
\ln \frac{\nu_s}{\nu_r} \approx \ln \frac{q_{\perp}}{Q q_{\perp}/m} = \ln \frac{m}{Q}\ .  
\ee
Similar to Eq.~\eqref{SHQR1}, the logarithms can be resummed as  
\be 
\label{SHQrt1} 
\tilde{\mc{S}}_{H/\mQ}^{rt} (z,\bl{b}, \mu)  = \Bigl(\frac{\nu_{cs}}{\nu_r}\Bigr)^{2a_{\Gamma} (\mu, 1/\bar{b})} 
\tilde{S}_{rt} (\bl{b},\mu,\nu_s) \tilde{S}_{H/\mQ} (z,\bl{b},\mu,\nu_r).
\ee
Finally, the $\mu$-evolution kernel between $\mu$ and $\mu_0~(\mu\gg \mu_0)$ to NLL is given by 
\be 
\ln U^{rt}_{\mc{S}} (\mu, \mu_0) = \ln \frac{m^2}{Q^2} a_{\Gamma} (\mu,\mu_0) 
- \frac{2C_F}{\beta_0} \ln \frac{\as (\mu)}{\as (\mu_0)}\ .
\ee

\section{Scale variations for scales involving $1/\bar{b}$}
\label{sec:scale-var}
\begin{figure}[h]
	\begin{center}
		\includegraphics[width=10cm]{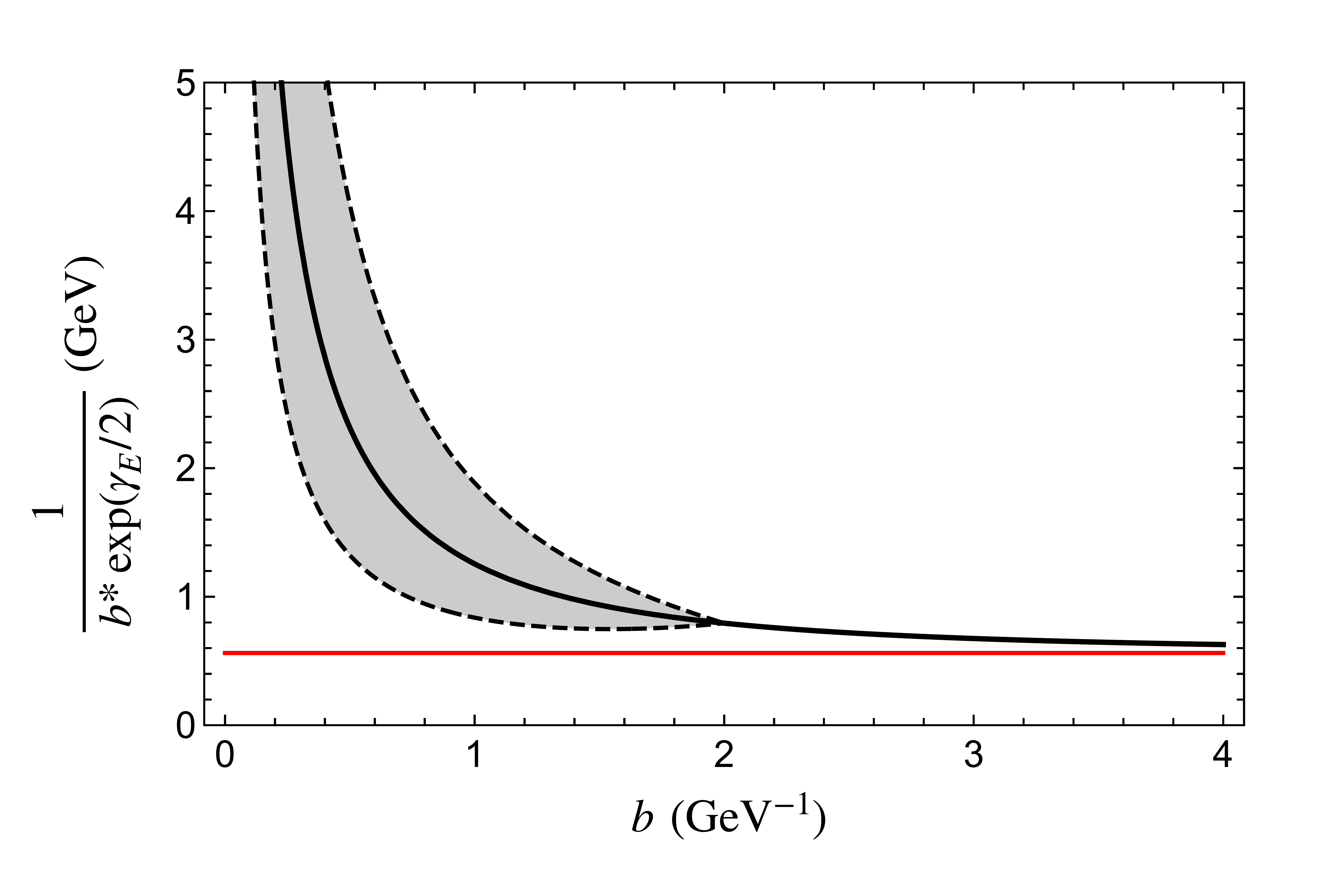}
	\end{center}
	\vspace{-0.5cm}
	\caption{\label{fig:scaleVar} scale variations for characteristic scales involving  $1/\bar{b}$. $b_{\rm max} = 2$ in the plot (and also in Figure~\ref{fig:pt}). The read line denotes the number  $1/(b_{\rm max} \exp(\gamma_E/2))$, and it is the non-perturbative scale which $1/\bar{b}$ is frozen into (for $b_{\rm max} = 2$, it is approximately equal to $0.56$ GeV).
	}
\end{figure}

For each characteristic scale involving $1/\bar{b}$, e.g. $\mu_c$ or $\mu_{cs}$ in Eq.~(\ref{eq:mu-scales}), we vary it according to what shows in Figure~\ref{fig:scaleVar} where we do what follows. First, $1/\bar{b}$ in the scale is replace with $1/(b^* \exp(\gamma_E/2))$, where $b^*$ is given in Eq.~(\ref{eq:bstar}) and $\gamma_E$ the Euler–Mascheroni constant. We then introduce a simple scaling function
\begin{equation}
	s(b) = 
	\begin{cases}
		2 - \frac{b}{b_{\rm max}} & \text { if } b < b_{\rm max},\\ 
		1 & \text { if } b \geq b_{\rm max}
	\end{cases}
\end{equation}
where $b_{\rm max}$ is the same as that appearing in defining $b^*$. Finally, the scale variation is carried out in the interval
\begin{equation}
	\left( \frac{1}{s(b) b^* \exp(\gamma_E/2)},  ~ \frac{s(b)}{ b^* \exp(\gamma_E/2)}\right).
\end{equation}

%%%%%%%%%%%%%%%%%%%%%%%%%%%%%%%%%%%%%%%%%%%%%%%%%%%%%%%%%%%%%%%%%%%%%%
%%%%%%%%%%%%%%%%%%%%%%%%%%%%% Bibliography %%%%%%%%%%%%%%%%%%%%%%%%%%%
%%%%%%%%%%%%%%%%%%%%%%%%%%%%%%%%%%%%%%%%%%%%%%%%%%%%%%%%%%%%%%%%%%%%%%

%%%%%%%%%%%%%%%%%%%%%%%%%%%%%%%%%%%%%%%%%%%%%%%%%%%%%%%%%%%%%%%%%%%%%%

%\phantomsection
%\addcontentsline{toc}{section}{References}

\bibliographystyle{JHEP1}
\bibliography{fullrefs}

%%%%%%%%%%%%%%%%%%%%%%%%%%%%%%%%%%%%%%%%%%%%%%%%%%%%%%%%%%%%%%%%%%%%%%

%%%%%%%%%%%%%%%%%%%%%%%%%%%%%%%%%%%%%%%%%%%%%%%%%%%%%%%%%%%%%%%%%%%%%%

\end{document}